%% file: Vortextransport-eps.tex
\documentclass[twocolumn,showpacs,preprintnumbers,amsmath,amssymb]{revtex4}
\usepackage{dcolumn}
\usepackage{color}
\usepackage{bm}
\usepackage{graphicx}
\input ldefs

\def \be{\begin{equation}}
\def \ee{\end{equation}}
\def \bea{\begin{eqnarray}}
\def \eea{\end{eqnarray}}

\def\bnabla{{\boldsymbol{\nabla} }}

\def\bap{\bfA\nd_\perp}

\physgreek

\begin{document}
\title{Vortex Tunneling and Transport Theory In Two-Dimensional Bose Condensates}
\author{Assa Auerbach$^{1}$, Daniel P. Arovas$^{2}$ and Sankalpa Ghosh$^{1,3}$}
\address{$^1$Physics Department, Technion, Haifa  32000, Israel\\
$^2$Department of Physics, University of California at San Diego, La Jolla, CA 92093\\
$^3$Physics Department,
Okayama University,
Okayama-700-8530, Japan}
\date{\today}
\begin{abstract}
The tunneling rate $t_v/\hbar$ of a vortex between two pinning sites (of strength $\bar{V}$ separated by $d$) is computed
using the Bogoliubov expansion of vortex wavefunctions overlap. For BCS vortices, tunneling is suppressed beyond
a few Fermi wavelengths.  For Bose condensates,
$t_v  = \bar{V} \exp({-\pi n_s d^2/2})$, where $n_s$ is the boson 
density. The analogy between vortex hopping in a superconducting film and 
2D electrons in a perpendicular magnetic field is exploited.  
We derive the variable range hopping temperature, below which vortex tunneling contributes to magneto-resistance.
Using the 'Quantum Hall Insulator'  analogy we argue that the {\em  Hall conductivity} (rather than the inverse Hall resistivity)  measures
the effective carrier  density in  domains of mobile vortices.

Details of vortex wavefunctions and overlap calculations, and 
a general derivation of the Magnus coefficient for any wavefunction on the sphere, are provided in appendices. 
\end{abstract}
\maketitle
\vskip2pc 
\narrowtext


\section{Introduction}
Mass and charge transport in  superfluids, superconductors and Bose Einstein Condensates (BEC) are  
governed by mobility of  vortices  \cite{and66,tinkham}.
In two dimensions, vortex centers effectively interact as point charges in a perpendicular magnetic (Magnus) field.
In the superfluid phase, vortices are pinned at zero temperature  by impurity potentials  \cite{AHNS80,tinkham}.  
Their mobility, just below the pinning temperature scale, is dominated by thermally activated hopping  \cite{AK64}, essentially
following a classical Arrhenius law.

At lower temperatures, quantum
fluctuations may, in principle, admit tunneling of vortices under energy barriers, resulting in  'quantum flux creep'  \cite{tunn-particle,eck89,non-linVI}. Experimentally,
vortex tunneling in superconducting films, manifests itself as low temperature magnetization relaxation  \cite{Creep-exp} and  non-activated, variable range hopping resistivity  \cite{mpaf91}. As magnetic field and disorder strength increase, vortex tunneling can turn into long range delocalization.
This amounts to a quantum phase transition from the superfluid into an insulating  \cite{mpaf90,SCI-old,SCI-new}, or  perhaps a Bose metal 
phase  \cite{BoseMetal}.

A microscopic   computation
of vortex tunneling rates, has been an elusive theoretical goal.
The semiclassical (instanton) approach  \cite{vol95},  requires the determination of the   'vortex mass' in the presence
of  Magnus dynamics, and low energy superfluid phonons  \cite{NAT94,duan,simanek,aro97}.

In the presence of short range, localized pinning potentials, it is simpler to compute  (as we show below) 
the vortex tunneling rate from the many-body  wave-function overlap  \cite{Sonin,NAT94,assaQH,mizel}.  

\begin{figure}[htb]
\begin{center}
\includegraphics[width=7cm,angle=0]{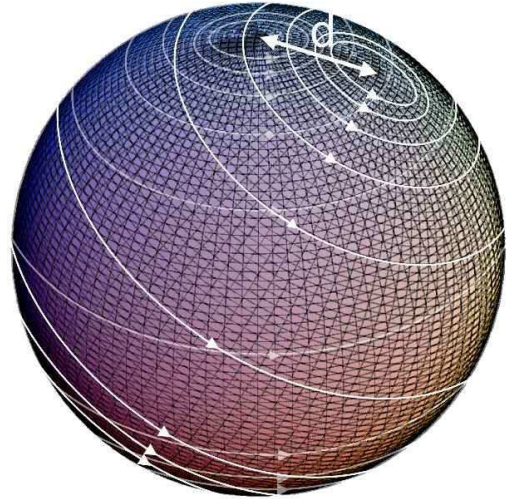}
\caption{Overlap of two antipodal vortex-antivortex pair states on a sphere, shifted by distance $d$,
which are used in this paper to numerically evaluate the vortex tunneling rate.  
White circles depict directions of circulating currents.} 
\label{fig:antipodal}
\end{center}
\end{figure}

In this paper we present  detailed vortex overlap calculations for the following systems:

(i) {\em The weakly interacting Bose Einstein Condensate (BEC)}.
The tunneling rate between two localized
pinning potentials of strength $\bar{V}$, separated by distance $d$,  is determined  to be
\be
t_v = \bar{V} \exp\left( -C {\pi\over 2} n_s d^2+ \cO(1/n_s\xi^2) \right),
\label{tv-intro}
\ee
where  $n_s$, and $\xi$ are the boson number density and coherence length respectively. 
Numerically $C= 1\pm 0.02$. The tunneling rate  (\ref{tv-intro}) applies to low density BEC's and their charged version, the  
'Bosonic superconductor'.

(ii) {\em The BCS superconductor}. Conventional superconductors, with large electron density and core radius, have a tunneling rate which
is   suppressed  by a factor of $e^{-0.2 k_F^2 d^2}$, where $k_F$ is
the Fermi wavevector. Thus, one can conclude that 
BCS vortices cannot observably tunnel under barriers larger than a few Fermi wavelengths.
 
Our results point to the experimental regime where vortex tunneling in superconductors may have measurable effects on magneto-transport.
Basically,  the superfluid density should be much lower than the usual metallic electron density.  This condition
may be  realized in  cuprate (High T$_c$) films  \cite{Creep-exp}, especially in the underdoped regime  \cite{Uemura}, 
and in highly disordered superconducting films  \cite{SCI-old,SCI-new}, where phase fluctuations are important.

The paper is organized as follows. 

 Section \ref{sec:tunn} sets up the tunneling calculation by expressing the two site vortex tunneling rate in terms of 
the pinning energy and ground states overlap. 

Section  \ref{sec:overlaps} presents the overlap calculations of the interacting BEC (arriving at Eq. (\ref{tv-intro})), and the BCS superconductor. 

Section \ref{sec:transport} presents the  vortex  transport theory and its relation to electrical conductivity in superconductors.
We use a quantum hopping Hamiltonian based on Eq.(\ref{tv-intro}) coupled to the QED field of superfluid phonons
  \cite{simanek,aro97}.
Following the theory of Ambegaokar, Halperin and Langer  \cite{AHL} (AHL), we 
derive the variable range hopping exponent and temperature scale. 
The analogy to the 'Quantum Hall Insulator'  \cite{QHI-theory,QHI-Shahar,SA} is utilized to argue that the Hall {\em  conductivity}
of a bosonic superconductor
is a robust measure of its boson number density $n_s$. (This is in contrast to
the commonly used  assignment of the 'Hall number' using the inverse Hall {\em resistivity}).

We provide pedagogically instructive details  in a series of appendices: the overlaps of 
vortex mean field state, in App. (\ref{app:OVC}), the Bogoliubov theory, in App. (\ref{app:Bog}), and Bogoliubov de-Gennes
equations, in App. (\ref{app:BdG}),
and their matrix formulations in the spherical geometry, in App. (\ref{app:calcs}).

In Appendix (\ref{app:Magnus}) we prove that the Magnus coefficient (adiabatic curvature) of an {\em arbitrary}  wavefunction
on the sphere is given by its average {\em angular momentum} density. This result, which is  peripherally connected to the
main subject of this paper,  generalizes a previous proof connecting the vortex Magnus action to the far field density
  \cite{TAN96}. 

We conclude in Section  \ref{sec:summary} with a summary and a discussion.

\section{Vortex Tunneling and Wavefunction Overlap}
\label{sec:tunn}
The relation between the vortex tunneling rate and  wavefunction
overlap  follows the method of non-orthogonal ground states first introduced by Heitler and London   \cite{HL}. 

Let us consider a homogeneous condensate described by 
a many-body interacting Hamiltonian $\cH_0$. 
We perturb the system with two weak,
and symmetrically situated, pinning sites 
 at positions $\bfx_1$ and at $\bfx_2$. The full Hamiltonian is

\be
\cH = \cH_0+ \cV_1+\cV_2,
\ee
where $\cV_i=\lambda\,n(\bfx_i)$.  The pinning potentials
are repulsive for the bosons ($\lambda>0$) and are therefore attractive for the vortices.
The separation between the pinning sites is $d=|\bfx_2-\bfx_1|$.  We assume for this exercise that a reflection symmetry exists in 
$\cH$ about a mirror plane between the pinning sites.
The ground states  $\ket{\rmPsi^0_i}$ of the two {\em partial} Hamiltonians satisfy
($\cH_0+\cV_i)\ket{\rmPsi^0_i}=E_0\ket{\rmPsi^0_i}$.  Thus we can use the symmetric and
antisymmetric superpositions as a variational ground states for the two symmetry sectors, 
\be
\ket{\rmPsi^0_{\pm}}={\ket{\rmPsi^0_1}\pm\ket{\rmPsi^0_2}
\over \sqrt{2(1\pm |\braket{\rmPsi^0_1}{\rmPsi^0_2}|) }}
\ee

Their corresponding energies are bounded by 
\be
E_{\pm}^0 \le  E_0 + {\expect{\rmPsi^0_1}{\cV_2}{\rmPsi^0_1}\pm \expect{\rmPsi^0_1}{\cV_2}{\rmPsi^0_2}\over
 1\pm |\braket{\rmPsi^0_1}{\rmPsi^0_2}| }
 \ee
The coherent tunnel splitting  is defined as  $t_v =(E^0_+ - E^0_-)/2$.
To first order in the overlap, $\braket{\rmPsi^0_1}{\rmPsi^0_2}$,  we obtain
(see comment   \cite{Comm-HL}):
\bea
t_v  &\simeq& \Big|\expect{\rmPsi^0_1}{\cV_2}{\rmPsi^0_1}\cdot
\braket{\rmPsi^0_1}{\rmPsi^0_2} - \expect{\rmPsi^0_1}{\cV_2}{\rmPsi^0_2}\Big|\nonumber\\
&\simeq &{\bar V}   \big|\braket{\rmPsi^0_1}{\rmPsi^0_2}  \big|\ ,
\label{epm}
\eea
where $\overline{V} \equiv\expect{\rmPsi^0_1}{\cV_2}{\rmPsi^0_1}\approx \lambda n_1(\bfx_2)$.
We neglect the term proportional to $\expect{\rmPsi^0_1}{\cV_2}{\rmPsi^0_2}$ since it depends
on the density at the vortex center, which is assumed here to be small.

Thus, Eq. (\ref{epm}) establishes that  the pinning potential  supplies the `attempt rate' of the tunneling. We shall see that
the vortex wavefunctions' overlap decreases  as a gaussian of their separation $d$.

\subsection{Overlap exponent and compressibility}
\label{sec:NAT}
We briefly review the important result of Niu, Ao and Thouless (NAT)   \cite{NAT94} concerning vortex overlap.
Consider a vortex centered at $\bfX$, with core radius $\xi$. The Onsager-Feynman  \cite{OF} wavefunction 
is constructed from the uniform ground state $\Psi_0$, and is therefore
asymptotically correct for coordinates $\bfx_i$ far from the core center.
\be
\Psi_{\bfX}  \simeq \prod_i  \left( \exp\left(i \phi(\bfx_i-\bfX) \right) f (|\bfx_i-\bfX|/\xi)\right) \Psi_{0}
\ee
where $f(y)\to 1$ at $y>>1$.
The overlap between two such wavefunctions, displaced by a distance $d$, is given by
\bea
\langle \Psi_{\bfX}| \Psi_{\bfX+\bfd}\rangle&=&e^{-W_c} \exp\left(-  {\pi \over 2} n_s d^2   \int_0^{\bar{k}}{ dk \over k}   S (\bfk)\right),\nonumber\\
S(\bfk )&=& {1\over n_s} \int d^2\bfx \langle\Psi_0| \delta n(0) \delta n(\bfx)|\Psi_0\rangle e^{i \bfk \cdot\bfx}, \nonumber\\
\label{nat}
\eea
where $W_c$ is the overlap of the core area. $\phi(\bfx)$ is the angle between $\bfx$ and the $x$ axis.
$n_s$ is the average density, $\delta n = n-n_s$, and  
$\bar{k}\approx 2\pi/\xi$ is core wavevector cut-off.

The structure factor $S(k)$ is bounded by Bogoliubov's inequality  \cite{PS}
\be
 S(k) \le  \sqrt{ F(k) \chi(k)}  .
 \label{ineq}
\ee

For interacting bosons of mass $m$ and velocity independent interactions, the equal-time correlator  $F$ is given by 
\be
F(\bfk)\equiv {\hbar^2\over N} \langle \left[n_\bfk ,[H, \delta n_{-\bfk}]\right]\rangle =   {\hbar^2 k^2 \over 2m}  .
\ee
The compressibility $\chi$ is  related to the sound velocity $c_s$ by 
\be
\lim_{k\to 0} \chi(k) \to {1\over 2 m c^2_s}.
\ee
Inequality (\ref{ineq})  ensures that for the interacting BEC, the momentum integration in (\ref{nat}) converges at low $k$,
and therefore the vortex overlap exponent is finite in the thermodynamic limit.

We point out in Appendices (\ref{app:OVC}) and (\ref{app:BdG}), that both the Gross-Pitaevskii (mean field) coherent state and the BCS vortex wavefunctions,  
suffer from a spurious overlap catastrophe, due to  their unphysical,
infinite compressibilities  \cite{Thanksto}.

\section{Vortex Overlap Calculations}
\label{sec:overlaps}
\subsection{The Interacting BEC}
We consider a two-dimensional BEC with short range interactions,
described by the second quantized Hamiltonian
\be 
\cH =\!\int\!\!d^2\!x \left\{  \psi^\dagger K(\bfA)\,  \psi
+ V(\bfx)\,\psi^\dagger\psi 
  +{g\over 2}\,\psi^\dagger\psi^\dagger \psi  \psi\right\}\ ,
  \label{GPmodel}
  \ee 
$\psi^\dagger(\bfx) $ creates a  boson of mass
$m$ and charge $q$ at position $\bfx$; $\mu$ is the chemical potential.  The
single particle potential $V(\bfx)$ includes confining and vortex pinning contributions.
The kinetic operator is
\be
K(\bfA)={1\over 2m}\Big( {\hbar\over i}\bnabla+{q\over c}\bfA\Big)^2,
\label{K}
\ee
and $\bfA$ is a vector potential.
In the case of bosons of charge $q$ in a magnetic field $\bfB=B\,
{\hat\bfz}$ and subject to a rotation $\bfomega=\omega_{\rm rot}\,{\hat\bfz}$, the
vector potential is $\bfA=(\half B+ mcq^{-1}\omega_{\rm rot}){\hat\bfz}\times\bfx$, and the
single particle potential is shifted by $\rmDelta V(\bfx)=(\omega_{\rm rot}/2c)
(qB+m\,c\,\omega_{\rm rot})\bfx^2$.

If the system is a uniform droplet of bulk density $n_s$, the chemical potential gets pinned at $\mu = g n_s$.   
The important parameters of the condensate are
the phonon velocity  $c_s=\sqrt{gn_s/m}$, and the coherence length is $\xi=\hbar/mc_s$.

We use boson coherent states (\ref{varphi-mf}) to set up a semiclassical expansion of the  partition function, 
\bea
\cZ&=& \int\!\!\cD[\varphi,\varphi^*]\, \exp\left\{  \int\limits_0^{\hbar\beta}\!\! d\tau\!\!
\int\! d^2\!x\, \big( i\varphi^*\partial_\tau\varphi -\cH[\varphi^*,\varphi]\big)\right\}\nonumber\\
&\approx&\exp\left(-\beta E_{MF} [{\widetilde\varphi}]\right)\!\int\!\!\cD[\eta,\eta^*]\,
e^{- \cL^{(2)} [\eta^*,\eta]+\cO(\eta^3)}
\label{Zbog}
\eea
The first exponential is the classical (mean field) energy, and the remaining path integral  is over the fluctuation field $\eta=\varphi-{\widetilde\varphi}$.

The classical  field ${\widetilde\varphi}$ minimizes the variational energy  
$\langle \varphi  | \cH |\varphi\rangle$. It solves the Gross-Pitaevskii (GP) equation   \cite{Gross1,Pit1},
\be
 \left( K(\bfA)+V-\mu+g\big|{\widetilde\varphi}(\bfx)\big|^2 \right){\widetilde\varphi}(\bfx)=0\ ,
\label{GP}
 \ee
With a weak pinning potential at the origin, and an external magnetic field or
rotation   \cite{Fetter}, a stable vortex solution
can be found, whose approximate analytic form is   \cite{book1},
\be
{\widetilde\varphi}\approx {\sqrt{n_0}\,r \,e^{i \phi}\over \sqrt{r^2+\xi^2}}  ,
\label{barphi}
\ee
where $(r,\phi)$ are the polar coordinates of the vortex center.
For numerical evaluation of the fluctuation spectrum, the trial solution
(\ref{barphi}) must be improved upon by iterating Eq. (\ref{GP}).


\subsection{BEC Vortex overlap}
\begin{figure}[htb]
\begin{center}
\includegraphics[height=8cm,width=9cm,angle=0]{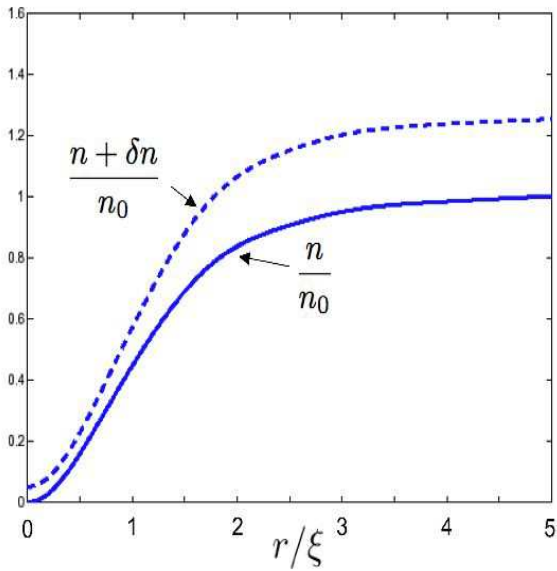}
\caption{The mean field condensate density profile $n_0(r)$ corrected by the Bogoliubov fluctuations density 
$\delta n_s(r)$ as a function  of radial distance $r$ from the vortex center. $n_0$ is the far field asymptotic density.}
\label{fig:deltan}
\end{center}
\end{figure}

In Appendix (\ref{app:OVC}) the mean field vortex coherent state is shown to suffer from an overlap catastrophe.
In Appendix (\ref{app:Bog}), the Bogoliubov-fluctuations corrected ground state is given by
\bea
 \big| \rmPhi  \rangle &=&\cN\exp\big(\half  \psi^\dagger\,Q\,\psi^\dagger\big)\,
\exp\big(f  \psi^\dagger\big)\, \big|0\big\rangle\nonumber\\
 f(\bfx)  & \equiv &  {\widetilde\varphi}(\bfx) -  \int\!\!d^2 x' \,{\widetilde\varphi}^*(\bfx') 
\, Q(\bfx',\bfx)  ,
\label{varphi-Bog}
\eea
where the operator $Q$ is determined by solving Bogoliubov's equations. 
The density profile of $|\Phi \rangle$ is
defined as
\be
\delta n(\bfx)=\langle \rmPhi   \big|  n(\bfx)-n_s \big| \rmPhi  \rangle ,
\ee
and plotted  in  Fig. (\ref{fig:deltan}) 
as  a  function  of radial direction from the vortex core. 
            
The BEC vortex wavefunction overlap is given by
\be
\big|\braket{\rmPhi_1}{\rmPhi_2}\big| = \exp\left( -W_0-W_1\right).
\label{W0W1}
\ee
We note that the leading order  $W_0$  is proportional to the condensate density $n_0$, 
while the fluctuation correction $W_1$ depends on $d,\xi, R$, but not on $n_0$. $W_1(R)$ turns out to diverge logarithmically with system size $R$, which is simply an artifact of  the Gaussian approximation in  $\big| \rmPhi  \rangle $. At this  sub-leading order, in $1/(n_0 \xi^2)$, the compressibility of this wavefunction diverges with $R$.
However, this need not concern us here, since we limit our calculation to the leading order which has a finite compressibility. 

\begin{figure}[htb]
\begin{center}
\includegraphics[width=9cm,angle=0]{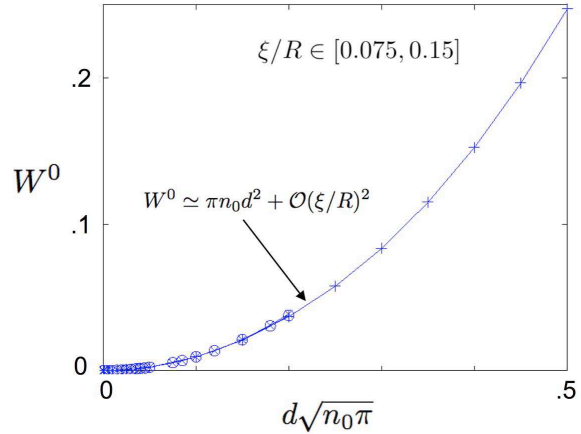}
\caption{The leading order overlap exponent for an antipodal vortex pair of a BEC on a sphere, displaced by a distance $d$. $n_0$ is the mean field condensate density.
Numerical data for  Eq. (\ref{unnorm-Bose-exp}) are marked by crosses and circles
representing different coherence lengths in the range $\xi/R \in[0.075.0.15]$.
The upper angular momenta cut-off was taken at $l_{\rm max}= 60$.  The solid line
represents the fit with $C=1$ in Eq.(\ref{bec-w0}).}
\label{fig:bec-w0}
\end{center}
\end{figure}
In Appendix (\ref{app:Bog}) we show that  $W_0$ of Eq. (\ref{sep}) is dominated by the core region.
A numerical calculation, described in Appendix (\ref{app:calcs}), is used determine $W(d,\xi)$.
A fit to the numerical
results for the antipodal vortex pair on a sphere of size $R$, depicted in Fig.(\ref{fig:antipodal}) yields,
\be
 W_0^{\rm sphere}(\bfd,R) = C \pi n_0 d^2 \big(1+\cO(\xi/R)\big),~~C= 1\pm 0.02.
 \label{bec-w0}
\ee

By halving the result for the vortex {\em pair} on a sphere (\ref{bec-w0}), an taking the large $R$ limit, we obtain the leading order result for a {\em single vortex} in the plane
\be
W_0 =\half \pi\,n_s d^2  +\cO(\xi/R, \delta n) ,
\label{W0-result}
\ee
which produces the result quoted  in Eq.(\ref{tv-intro}).

\subsection{BCS Vortex overlap}

In Appendix (\ref{app:BdG}) the vortex wavefunction overlap for an $s$-wave BCS superconductor 
is calculated  by numerically solving Bogoliubov-de Gennes equations. Analysis of the far field contributions, combined with the numerical results yields
\bea
\big|\braket{\rmPhi_1}{\rmPhi_2}\big| &=&  \exp\left( -W^{\rm BCS}\right),\nonumber\\
W^{\rm BCS}  &\simeq &  d^2 \left( {k_F\over 8\xi}\ln(R/\xi) +   k_F^2 F(k_F\xi)\right) ,
\label{WBCS1}
\eea
where $k_F$ is the Fermi wavevector, $\xi$ is Pippard's coherence length and the vortex core size,
and $F$ is a dimensionless function of magnitude ~0.2. The overlap catastrophe
(logarithm of system size $R$) comes from the far field contributions, and is an artifact of the infinite compressibility
of the BCS wavefunction. The important lesson
from (\ref{WBCS1}) is derived from the {\em core region} contribution  $e^{-0.2 k_F^2 d^2}$. This suppression
is not affected by the long wavelength phase fluctuations, but arises from the large density of electrons in the Fermi sphere which change their momenta when the vortex moves.
We can safely conclude therefore,  that in high superfluid density superconductors, 
vortex tunneling cannot  be observed beyond a distance of a few Fermi wavelengths.

 \section{Vortex Transport Theory}
 \label{sec:transport}

Mobile vortices imply the destruction of the static superconducting order parameter. However when order parameter (phase) 
correlations are of sizeable range,   it is  still convenient to describe 
the dissipative and Hall transport in terms of  dilute vortices,  rather than in terms of  a dense system of interacting bosons.
In two dimensional BEC's, the neutral  and charged  BEC have analogous transport equations, since shielding can be ignored.
The number current, pressure and rotation frequency in the neutral BEC play the role of electrical current, voltage and perpendicular magnetic field respectively, in the bosonic superconductor. For convenience, we shall discuss the latter case, keeping a keen eye on possible experimental ramifications.

\subsection{Vortex conductivity and electrical resistivity}
Steady dissipative vortex motion can be driven by a bias  current.  For bosons with charge $q$ the
current is $\bfj_c=qn_s \bfv_s $, where $\bfv_s$ is the superfluid velocity. 
The vortex 'charge' is  the sign of its vorticity, 
$Q_v= \pm 1$, and the vortex 'flux quantum' is unit of boson particle number $\Phi^0_v =1$.
The  kinetic energy difference  a vortex at site  $\bfR_i$ and $\bfR_j$ is
\be
E_0(i)-E_0(j)=   {h Q_v \over q}\bfj_c\times{\hat\bfz}\cdot(\bfR_i-\bfR_j).
\label{Ebias}
\ee
The  effective 'electric'  (actually {\em Magnus}) field acting on that  vortex  is 
\be
\bfve_v=  
{h \over   q } \bfj_c \times{\hat\bfz} .
 \label{bfEv}
 \ee
The stable ground state of a charged superfluid (i.e. a superconductor) in a magnetic field $B$ has a finite  
density of vortices given by
\be
 n_v  = B q/(hc),
\label{nv}
\ee
where $c$ is the speed of light. If the vortices have an average drift velocity $\bfV_v$ the vortex current is given by
\be
 {\cal \bfJ}_v =  {B q  Q_v  \over hc}\bfV_v .
\label{Jv}
\ee
The  electromotive  field (EMF) induced by the average vortex drift velocity is
\be
\bfE=-c^{-1}  \bfV_v \times\bfB = - {h \over q} \bf{\cal J}_v\times\hat{\bfz},
\label{lorentz}
\ee
where $c$ is the speed of light.

The 
{\em 'vortex conductivity'} tensor is defined as $\sigma_v$, 
\be 
{ \bf{\cal J}}_v^\alpha = \sum_\beta \sigma_v^{\alpha\beta }  \bfve_v^\beta .
\ee
Using ( \ref{bfEv}) and (\ref{Jv}),  the electric resistivity tensor 
$\rho$ is directly related to the vortex conductivity by duality relations \cite{mpaf90}
\bea
 \rho^{xx} &=&   \left( {h \over q }\right)^2 \sigma_v^{yy} ,~~(\&~~ y\to x),  \nonumber\\
 \rho^{xy}&=&  -\left( {h \over q }\right)^2 \sigma_v^{yx}.  \nonumber\\
 \label{rhoxx}
 \eea
 
 For example, consider unpinned vortices which can move in a Galilean invariant superfluid. 
 The vortex Hall conductivity 
 is given by analogy to a charged liquid of filling fraction $\nu$, which is $\sigma^{xy}=\nu e^2/h $, 
Setting $e\to Q_v$, one obtains
 \be
 \sigma_v^{xy} =  {Q_v\over h} n_v/n_s,
 \label{sigmav-class}
 \ee
 which by  using  (\ref{nv}) and (\ref{rhoxx}) yields (unsurprisingly), the classical  electric Hall resistivity of a  charged liquid with density $n_s$:
 \be
\rho_{\rm class}^{xy} = - {B\over n_s q c }.
\label{rhoxy-class}
\ee
This result will be used later, in Section (\ref{sec:Hall}).

\subsection{Vortex hopping hamiltonian}
Thus we arrive at  non-interacting vortex hopping hamiltonian is described by an Anderson tight binding model in a strong magnetic field
\be
\cH^{\rm v}=\sum_i \epsilon_i c\yd_i c_i  + \sum_{ij} t_v(d_{ij}) \left(e^{{i\over \hbar}\int_{\bfx_i}^{\bfx_j} \!d\bfx\cdot\bfa } c\yd_i c_j +{\rm h.c.} \right)
\label{Hv}
\ee
where $c\yd_i$ creates a vortex at a random pinning site $\bfx_i$ with random energy $\epsilon_i  $.
$\bfa$ is the Magnus gauge field which satisfies  
\be
\nabla \times \bfa = h n_s\hat{\bfz}
\ee
which gives  rise to a   Hall  effect.
The vortex hopping rate $t_v$ was precisely defined by (\ref{epm}). For a weakly interacting BEC it was
shown to decay with inter-site separation $d$ as a Gaussian
\be
t_v(d) \sim  \bar{V} \exp( -{\pi\over 2} n_s d^2 ).
\label{tv}
\ee

The low energy  vortex current operator is
 \be
 {\cal \bfJ}^v(\bfx)= {-i\over 2 \hbar} \sum_{ij} {\bf d}_{ij}  t_v(d_{ij}) \left( c\yd_j  c\nd_i -{\rm h.c.} \right) \delta\left(
 \bfx-\bfx_{ij} \right),
 \label{Jv1}
 \ee
 where $\bfd_{ij}$ and $\bfx_{ij}$ are the separation and midpoints of the pinning sites respectively.

Vortex dynamics have been systematically derived  for the two dimensional translationally invariant superfluid 
using an effective  Quantum Electrodynamics (QED) theory \cite{simanek,aro97}. In the QED formulation, vortices are point 'charges', moving in the presence of a 'transverse magnetic field' (the condensate density), and interacting with 'photons' (the Bogoliubov phonons).
The photons give rise to a vortex self energy, which diverges  logarithmically at low frequencies. This diverging 'effective mass'
however does not preclude quantum tunneling at finite  timescales \cite{timescale}. 

Following the derivation of Ref. \cite{aro97}, we explicitly retain the low energy phonons by  coupling them as a gauge field to the vortex current:

\bea
\cH^{\rm v-ph}&=&  \int d^2 x  {\cal \bfJ}_v(\bfx )  \cdot   \bap(\bfx )+ \sum_\bfk \hbar c_s |\bfk| a\yd_{\bfk} a\nd_{\bfk} ,\nonumber\\
\bap&=& {i h\over \sqrt{V}}\sum_\bfk e^{i\bfk\cdot \bfx }  \bigg({ n_s \xi  \over 2  |\bfk| }\bigg)^{1/2}
\hat{\bfz} \times{\hat\bfk}\,\big(a\nd_\bfk+a\yd_{-\bfk}\big) ,   \nonumber\\
\label{v-ph}
\eea
where  $a^\dagger_\bfk$ creates a Bogoliubov phonon ('photon' in the QED language \cite{aro97}) of 
wavevector $\bfk$ and frequency $\hbar c_s \bfk$, and $c_s=\hbar/(m\xi)$ is the speed of sound.
The vortices are treated as hard-core particles, which like adsorbates on a surface, have a Fermi-Dirac occupation probability 
\be
n_i = \left(e^{(\epsilon_i-\mu)/T}+1\right)^{-1}.
\ee
We  set the chemical potential $\mu$ to zero, and  fix the  average density $n_v=\sum_i n_i /N_{\rm pin}$
by the magnetic field as given by  (\ref{nv}). 
By inserting (\ref{Jv1}) in Eq. ( \ref{v-ph})
for two sites separated  by $R$, the two site conductance is given by the hopping theory  ( \cite{hopping})
\be
G^v(d) 
=\gamma_0  e^{-\pi n_s d^2 } e^{- {1\over 2T}\left(|\epsilon_i| + |\epsilon_j| + |\epsilon_i-\epsilon_j|\right)}
\label{Gv}
\ee
where  and
\bea
\gamma_0&=&{2\pi   n_v  \bar{V}^2  d^4 \over \hbar^4 T} \cR^{ph} (\epsilon_i-\epsilon_j) , \nonumber\\
\cR^{\rm ph} (\omega)&=& \Im \int_\infty^\infty\!  dt   e^{-i\omega t} \langle \bap(\bfx,t)\bap(\bfx,0)\rangle\nonumber\\
 &=& {h^2 n_s \xi \over 4  c_s }(1+N_b(\omega/T)) .
 \label{gamma0}
 \eea
$\cR^{\rm ph}$ is the bosons local dissipative response to the vortex motion, and $N_b$ is the Bose function.

 \subsection{Vortex Variable Range Hopping}
 \label{sec:VRH}
The {\em macroscopic} vortex conductivity for the hopping hamiltonian (\ref{Hv}) requires knowledge of
the distribution of  pinning site positions and energies $\cP(\{\bfx_i,\epsilon_i\})$. 
We focus our attention on {\em individual} vortex tunneling, in the regime of
low vortex densities
$n_v<<n_{\rm pin}$ and small  random fluctuations $\delta\bar{V}<<\bar{V}$. Interaction effects between
vortices are self consistently incorporated into  $\epsilon_i$. 

The density of states depends on both the pinning potential distribution, and the effects of vortex interactions. 
Here, we shall  treat the low field, finite temperature regime where the mean hopping distance is much smaller than the vortex separation.
The  density of states is then reasonably modelled by
\be
\cN(\bfx,\epsilon)=\langle \sum_i\delta(\bfx-\bfx_i)\delta(\epsilon-\epsilon_i)\rangle = {n_{\rm pin} \over\delta\bar{V}} ,
\label{distribution}
\ee
where  $n_{\rm pin}$ is the  pinning site density, and its energies are uniformly distributed 
in the interval $(-\delta\bar{V}/2, \delta\bar{V}/2)$.

The vortex conductivity maps onto 
Mott's variable range hopping (VRH) \cite{Mott}  of charges in a random potential and a strong magnetic field.
The gaussian decay of Eq. (\ref{tv}) is directly analogous to that of an electron in the lowest Landau level  \cite{vrh-Sh,Fogler},   where the  'Landau length' is  $\lambda= \sqrt{2\pi n_s}$.

At low enough temperatures $T<\delta\bar{V}$,  there are typically many  
competing tunneling paths between pinning sites separated by  distances $d >>\lambda, 1/\sqrt{n_{\rm pin}}$.
As Shklovskii has shown \cite{vrh-Sh,Fogler}, in this regime  multiple (virtual)  tunneling processes play a crucial role. Their primary effect
is to replace the gaussian decay of the two-site tunneling rate  by a linear
decay, typical of an Anderson insulator. Here,  we must therefore replace $t_v(R)$ of (\ref{tv}) by 
\be
\tilde{t}_v \simeq \bar{V} e^{-d/\ell}, ~~\ell={  2 s \sqrt{n_{\rm pin}}\over \pi n_s} ,
\label{tilde-tv}
\ee
where $\ell$ is the linear localization length, and $s$ is a numerical which depends on the details of $\cP(\{\bfx,\epsilon\})$.
Replacing (\ref{tv}) by (\ref{tilde-tv}),  we arrive at a two dimensional random resistor network, of the kind discussed by  
 Ambegaokar, Halperin and Langer \cite{AHL} (AHL), with random
conductances  given by
 \be
 \tilde{G}_v^{ij} = \gamma_0 \exp\left( -{2 d_{ij}\over \ell} - {|\epsilon_i| +| \epsilon_j| + |\epsilon_i-\epsilon_j|\over 2T} \right) .
 \ee

By AHL, the macroscopic conductance is given by the critical (lowest) conductance $G_v^c$ of the percolating subset
of conductances which obey $\tilde{G}_{ij}\ge G_v^c$.

 Taking the average number of bonds per site at percolation
 to be $\nu_c$ (e.g. on the square lattice $\nu_c=2$), the percolating bonds all obey
  \be
{ 2d_{ij}\over\ell}+   {|\epsilon_i| +| \epsilon_j| + |\epsilon_i-\epsilon_j| \over 2T} \le  \ln(\gamma_0/G_v^c),
 \ee
 which can be written as
\be
{ d_{ij}\over d_{\rm max} }+   {|\epsilon_i| +| \epsilon_j| + |\epsilon_i-\epsilon_j| \over 2\epsilon_{\rm max}} <1,
 \ee
 where 
 \bea
d_{\rm max} &=&  {\ell\over 2}    \ln(\gamma_0/G_v^c) , \nonumber\\
 \epsilon_{\rm max} &=&  T  \ln(\gamma_0/G_v^c) .
 \label{dmax}
 \eea
 
 By (\ref{distribution}) the density $n_{\rm conn}$  of connected sites 
 within $|\epsilon_i|\le  \epsilon_{\rm max}$ is given by
 \be
 n_{\rm conn}= n_{\rm pin} { \epsilon_{\rm max}  \over \delta\bar{V}} .
 \label{nconn}
 \ee
The percolation condition on the number of connections per site is
 \be
 n_{\rm conn} d_{\rm max}^2 =\nu_c ,
 \label{nuc}
 \ee
which implies the relation
 \be
 \epsilon_{\rm max}  d_{\rm max}^2={ \nu_c \delta\bar{V}  \over n_{\rm pin}}.
 \label{relation}
 \ee
 Using (\ref{dmax}) and (\ref{relation}) one obtains the value of the critical vortex conductance
  \bea
G_v^c &=& \gamma_0e^{ -\left({T_0/T}\right)^{1\over 3}} ,\nonumber\\
 T_0 &=& K \delta\bar{V} \left({\pi n_s\over n_{\rm pin}}\right)^2 ,
 \label{Gc}
 \eea
 where $K= 4 \nu_c/s^2$ is a dimensionless  factor of order unity.  Using (\ref{rhoxx}) 
 we obtain the variable range hopping magnetoresistivity 
\be
\rho^{xx}(B,T) =  \left({ h\over q}\right)^2\gamma_0(n_v(B)) e^{ -\left({T_0/ T}\right)^{1\over 3}}.
 \label{RT}
 \ee

$\rho^{xx}(B,T)$  exhibits vortex tunneling in two ways. First, the   power of $1/3$
in the exponential temperature dependence. 
Second, the  VRH temperature scale $T_0$ depends strongly on the ratio of  vortex tunneling lengthscales:
the characteristic tunneling distance $1/\sqrt{n_{\rm pin}}$  divided by
the inter-boson separation $1/\sqrt{n_s}$.

At stronger fields (higher vortex density),  long range vortex interactions are expected to
modify the asymptotic power of the hopping exponent \cite{mpaf91}.

\subsection{Vortex Hall Resistivity}
\label{sec:Hall}
\subsubsection{Quantum Hall Insulator: review}
Early on, Holstein  \cite{Hol-Hall} has studied the  Hall effect of the hopping model (\ref{Hv}) at low temperatures.
He has shown the  importance of three site tunneling interference, for
producing a non zero Hall effect.

Since then, several groups have extended that work to electrons in highly disordered two dimensional
semiconductors in the presence of a strong magnetic field \cite{QHI-theory}. Although
different approximation schemes were used, these groups have concluded that 
while $\rho^{xx}$ diverges at low frequency and temperature, $\lim_{T\to 0,\omega\to 0}\rho^{xy}(\omega,T) <\infty$.
Such behavior was dubbed 'Quantum Hall Insulator'  (See comment  \cite{sxy-omega}).

Experiments in Hall bars \cite{QHI-exp}, have found that the DC Hall resistivity has a much weaker  temperature dependence than
the resistivity on the insulator side of  the field tuned 
metal-insulator transition. Refs. \cite{QHI-Shahar} have remarkably found the Hall resistance
in the insulator to be  {\em quantized} at $h/e^2 \nu$, at filling factor of $\nu=1/3$.
 
The apparent difference between the behavior of the Hall resistance versus the longitudinal resistance, can be explained by  Kirchoff's transport theory
of an inhomogeneous resistor network \cite{Hall-class}, with widely varying resistances. The  Puddles Network Model
(PNM)  \cite{SA}, was introduced to explain the experiments of Refs. \cite{QHI-Shahar}.
The PNM assumes a network of perfect Hall liquid puddles with conductivities $\sigma^{xy}=\nu e^2/h$, $\sigma^{xx}=0$,
embedded in an insulating environment and connected by arbitrary large, classsical resistors. This model
yields a quantized value of $\rho^{xy}=h/e^2 \nu$, independent of $\rho^{xx}$.

While the PNM describes ohmic (incoherent) transport,   quantum transport theory yields a different result.
Using the  Chalker-Coddington network    
to represents  non interacting electrons in the lowest Landau level in the presence of smooth
disorder,  Ref.  \cite{PA}
has numerically found that the Hall resistance at zero temperature actually diverges with system size,
similarly to  the quantum induced localization of $\rho^{xx}$. Therefore, a true Hall insulator phase for the Chalker-Coddington model has been ruled out. The conflicting results of classical and quantum transport theories, is related to the 
role of dephasing. Inelastic scattering  destroys localization, and prevents the divergence of $\rho^{xy}$.

\subsubsection{Quantum Vortex Insulator}
The above discussion is directly relevant to the vortex  hopping model (\ref{Hv}). The vortices are essentially in an insulating state with possible
domains of weaker superconductivity where vortex mobility is higher. A diverging $\sigma^{xy}$ at low temperatures
may indicate long range coherent  vortex transport, an interesting result in itself.

Let us for now assume sufficient dephasing at the low temperature of experiments, due to vortex-phonon, or vortex fermion interactions.
We can appeal to the Boltzmann transport theory, and to the resistor network models. 
This implies that the Hall {\em conductance} (not resistance) is determined
by the Hall conductivity $\sigma^{xy}$ of the most 'insulating' puddles. We do not know how to compute the distribution of $\sigma^{xy}$.
However,  by  (\ref{sigmav-class}),  $\sigma^{xy}$ measures the effective carrier density $n^*$ in the most resistive domains,
(i.e. the vortex liquid puddles):
\be
\sigma^{xy}(B)  \sim  n^*_s q c/B
\label{sxy}
\ee
Furthermore, a detection of  'quantized' plateaux of $\sigma^{xy}(B,n_s)$ may indicate  locked-in  charge density waves or
topological ordering \cite{Balents} in the  vortex-condensed domains.

\section{Experimental Implications and Discussion}
\label{sec:summary}
{\em Cold atoms BEC}. Vortices have been created in rotating cold atomic gases \cite{vortex-bec}. One can imagine optically introducing localized pinning potentials
and measuring the excitation spectrum. The lowest antisymmetric excitation  could be compared to expression (\ref{tv-intro}), for different potential separations and boson densities.

{\em Cuprate superconductors}. In thin cuprate films, time resolved magnetization relaxation \cite{Creep-exp}, is a direct measure of the average vortex mobility.
A variable range hopping behavior of the magneto-resistance section is indicative of tunneling effects, as was shown in section (\ref{sec:VRH}).
For a 'bosonic superconductor' (coupled only to order parameter phase fluctuations), the characteristic resistivity given in  Eq. (\ref{RT}) is   

\be
{h^2 \over q^2 } \gamma_0 \sim {h \over q^2} \left( {n_v\over n_{\rm pin}} \right) \left( {\bar{V}\over T} \right)
\left( {n_s\over n_{\rm pin} } \right) n_s \xi^2 .
\ee
This expression has dubious applicability to high T$_{\rm c}$ films. However, if one accepts a model of
tighly bound hole pairs, with a low superfluid density, the vortices  can primarily dissipate momentum 
to the low energy 'nodal' fermions, and the core states near the vortex center\cite{QED3,PBFM}. 
Incorporating fermionic excitations amounts to adding  a dissipative response to $\cR^{\rm ph}$,  given by
\be
 \cR^{\rm fer}(\omega) = \left({ h^2 c^2 \over e^2 \omega}\right) \sigma^{\rm ferm} (\omega).
\label{Rferm}
\ee
where $\sigma^{\rm ferm}(\omega)$ is the fermions contribution to the AC conductivity. Crudely estimating
the  factors contributing to
the characteristic resistivity, we obtain
\be
{h^2 \over q^2 } \gamma_0 \sim {h \over q^2} \left( {n_v\over n_{\rm pin}} \right) \left( {\bar{V}\over T} \right)
\left({\bar{V} \over \delta\bar{V}} \right) \left( { \sigma(\delta\bar{V}/\hbar)\over e^2 /h}\right)
\ee
The values of $\bar{V}$ and $\delta\bar{V}$ may be extracted from the resistance activation energy
at higher temperatures. 

{\em Hall Conductivity}. Hall effect measurements in underdoped cuprates, have determined the Hall number as
$n_H(T) = -B/(\rho^{xy} e c)$ and found it to be of the same sign and magnitude as the
{\em hole} doping concentration away from the Mott insulator phase  \cite{hall-hitc}. However, 
the 'anomalous' strong  temperature dependence of $n_H(T)$ has been used to distinguish the 
unconventional nature of the cuprates
which differs from the much weaker temperature dependence of the Hall number in
conventional metals.

Our analysis, suggests that  the Hall conductivity, Eq.(\ref{sxy}), rather than Hall resistivity
should be used to define a Hall number in the superconducting phase. 
In this regime, accessible by strong magnetic
fields \cite{bobinger},  $\sigma^{xy}$  is expected to be less temperature dependent and to characterize the Hall
coefficient of  metallic
'puddles'  inside the superconductor, where vortices are locally delocalized by  tunneling.

{\em Disordered Superconducting films}.  Highly disordered superconducting films \cite{SCI-old} are also
likely candidates for observing  vortex tunneling, since they effectively exhibit low superfluid density.
We  expect variable range hopping and a finite Hall conductivity
near the supercondcutor-insulator transition, where vortices become delocalized. However,  
we refrain from  quantitative estimates for these effects
since a microscopic theory for strongly inhomogeneous interacting fermion systems is beyond the scope of this paper.

 {\em  Periodic lattices}. Optical lattices of cold bosons and Josephson junction arrays introduce  the challenge of a strong periodic potential. The vortex hopping hamiltonian
(\ref{Hv}) can describe a  periodic lattice of weak pinning  potentials. One expects the lattice constant to play an important role in vortex mobility
as it does in the Hofstadter problem of a tight binding electron motion in a strong magnetic field. Indeed, 
recent theoretical work has shown \cite{Balents} that ground states degeneracies and vortex dynamics, 
depend on the boson filling per lattice site.

\section{Acknowledgements}
We thank Ehud Altman  and Anatoli Polkovnikov for correcting an erroneous conclusion in an earlier version of this work \cite{Thanksto}.
Conversations with S. Gayen, M. Fogler, A. Kapitulnik, A. Keren, S. Kivelson, D.-H. Lee, A. Mizel, E. Shimshoni, E. Sonin,
Z. Tesanovic, and O. Vafek are gratefully acknowledged.
We  acknowledge support from the US-Israel Binational Science Foundation and the fund for promotion of research at Technion.
The hospitality of Aspen Center for Physics, Kavli Institute for Theoretical Physics 
 and Technion Institute for Theoretical Physics, where portions of this research were carried out.

\appendix
 \section{Spurious Overlap Catastrophe in Mean field Theory}
 \label{app:OVC}
This appendix provides pedagogical examples which reflect  the limitations of  mean field theory of superfluids of superconductors. 
In particular,  we show  that Bose coherent states and Bogoliubov de-Gennes wavefunctions  have logarithmically divergent overlap exponents
in the thermodynamic limit, as a consequence of their  infinite compressibilities.

\subsection{Bose Coherent states}
Coherent states wavefunctions are often  used as  zeroth order approximations to ground states of Bosons,  
superconductors, and  quantum spin models with long range order. Here we show that
when describing a vortex, coherent states generically exhibit  an overlap catastrophe, i.e. a logarithmically diverging exponent.
This divergence is an artifact of the unphysical infinite compressibility exhibited by non-interacting bosons.

 Consider the coherent state
$\ket{\varphi}$, with $\varphi=\sqrt{n_0}\,e^{i\phi}$ a complex
scalar field, defined by
\be
\ket{\varphi}=
\exp\int\!d^2\!x \Big(\varphi(\bfx)\, \psi^\dagger(\bfx) -\varphi^*(\bfx)\, \psi(\bfx)\Big)
\ket{0}\ .
\label{varphi-mf}
\ee
\begin{figure}[htb]
\begin{center}
\includegraphics[width=7cm,angle=0]{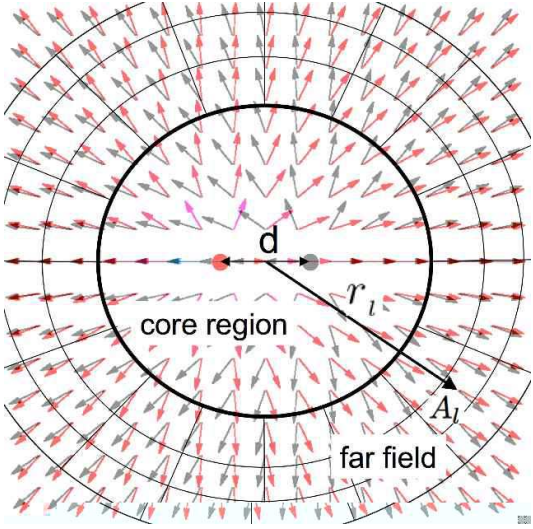}
\caption{Overlap of two vortex coherent states as described by Eq. (\ref{overlap-mf}).
Black (red) arrows are vector representation of $\varphi_1$ ($\varphi_2$).
For asymptotic calculations of BEC and BCS cases, see Eqs.(\ref{sep},\ref{ov-fer-far}),
 the far field region is divided into uniform phase blocks labeled by $l$ at distances $r_l$ and  areas $A_l$.}
  \label{fig:vor-cs}
\end{center}
\end{figure}
The overlap exponent between two  translated vortex coherent states (\ref{varphi-mf})), 
centered at  $\bfX=\pm \half\bfd$ is  then $|\braket{\varphi_1}{\varphi_2}|=
\exp(-W)$, with
\be 
W^{\rm CS} =  \int \!d^2\! x\,\Big[ \half \big(| {\varphi}_1|^2+ 
| {\varphi}_2|^2\big) -{\rm Re}\varphi_1^*\, {\varphi}_2  \Big]\ . \nonumber\\
\label{overlap-mf}
\ee 
The far field integral (away from the core), where
$n_i(\bfx) \simeq n_0$.  In this case, $\varphi\sim\sqrt{n_0}\,\exp(i\phi)$, where
$\phi$ is the angle function relative to the vortex center (see Fig. \ref{fig:vor-cs}).
\bea
W^{\rm CS}_{\rm far}(d) &=&   n_0\int\limits_{r_0}^R\! d^2\!x\, \Big[1-\cos(\phi_1-\phi_2)
\Big]\nonumber\\
&\approx&\half n_0\int\limits_{r_0}^R\! d^2x   \left({\bfd\cdot\bnabla\phi\over r}\right)^2 \nonumber\\
&=&\half\pi\,n_0 d^2  \ln(R/r_0)   ,
\label{MF-IR}
\eea
where $r_0$ is a near field cut-off. This calculation
was first presented by Sonin  \cite{Sonin}, in the context of vortex tunneling
using a noninteracting (product) condensate wavefunction.

For the non interacting Bose condensate wavefunction, it is easy to verify that the structure factor is  
\be
\langle \delta n(\bfx) \delta n(\bfx')\rangle = \delta(\bfx-\bfx') n_0
\ee
which yields finite zero momentum density fluctuations  $S^{\rm CS}(k) =1$. This is an artifact of the
unphysical limi t of {\em non interacting}  bosons, where the sound velocity vanishes and  the compressibility is infinite. 
The wavefunction overlap integral in (\ref{nat}) diverges logarithmically with the lower momentum cut-off, signaling this orthogonality catastrophe seen in 
(\ref{MF-IR}). Recall, however,  that non interacting bosons do not support stable vortices
as their coherence length is infinite!

 \section{Bogoliubov fluctuations}
\label{app:Bog}
The fluctuations  $\eta$ in (\ref{Zbog}), are  governed by the harmonic action
\be 
S^{(2)}= \half \int\!d\tau\!\int \! d^2\!x\,
\begin{pmatrix}\eta^* & \eta\end{pmatrix}\,\left[\,  iJ\partial_\tau- H^{(2)}\right]
\begin{pmatrix} \eta \\ \eta^* \end{pmatrix},
\ee
where
\bea
 J&\equiv& \begin{pmatrix}  1&0 \\
0& -1\end{pmatrix}\quad,\quad
H^{(2)}=   \begin{pmatrix} H_0  &&g{\widetilde\varphi}^2 \\ &&\\
g {{\widetilde\varphi}^*}{}^2&& H_0^*\end{pmatrix},
\label{HBog}
\eea
and $H_0=K(\bfA)+2g\,|{\widetilde\varphi}(\bfx)|^2-\mu$.

The Hamiltonian $H^{(2)}$ is diagonalized by the canonical transformation,
\be 
S= \begin{pmatrix}U   &V^* \\
V& U^*\end{pmatrix} \quad, \quad S^\dagger
 H^{(2)}   S
= \begin{pmatrix}  \hat{E}   &0 \\
0& \hat{E}    \end{pmatrix},
\ee
with
\be 
 S^\dagger \hat{J} S= \hat{J}=S\hat{J} S^\dagger \ .
 \ee
Here $\hat{E} =\delta_{nn'}E_n$ is the Bogoliubov spectrum  \cite{Burnett}.   The spectrum and
eigenoperators are explicitly determined  by solving the differential equations,
\bea
H_0\, U_n (\bfx) +  g {\widetilde\varphi}^2\,  V_n  (\bfx)   &=&
+E_n  U_n (\bfx) \nonumber\\
g{\widetilde\varphi}^*{}^2\, U_n(\bfx) +H_0^* \, V_n(\bfx) &=&  -E_n  V_n(\bfx),
\label{BoseBdG}
\eea
\subsection{Bogoliubov corrections in the far field approximation}
\label{App:unifBEC}
In a large area, asymptotically far away from the vortex core, we can solve the Bogoliubov equations (\ref{BoseBdG}) using a constant order parameter, 
${\widetilde\varphi}=\sqrt{n_0}\,e^{i\phi}$.
$H^{(2)}$  is diagonal in Fourier space, and the matrix $Q$ is given by
\be
Q_{\bfk,\bfk'} = -{e^{2i\phi}\over\mu}\big( \ve_\bfk-\mu-E(\ve_\bfk)  \big) 
\delta_{\bfk,\bfk'}\equiv   e^{2i\phi-\theta_\bfk}
\label{uniform}
\ee
where $\ve_\bfk=\hbar^2\bfk^2/2m$ and $E(\ve)=\sqrt{\ve^2+2\mu\ve}$.
The Bogoliubov fluctuations lead to an increase in the density. Defining
\bea
\delta n &=&  \expect{\rmPhi }{\eta\yd\eta}{\rmPhi }= \half \sum_{\bfk} v^2(\ve_\bfk) \nonumber\\
v^2(\epsilon)&=&{\mu^2\over 2 \sqrt{(\epsilon+\mu)^2-\mu^2}\left(\epsilon+\mu+ \sqrt{\epsilon+\mu)^2-\mu^2}\right)}. \nonumber\\
\eea
Changing variables to $y=1+\epsilon/\mu$, one obtains
\bea
\delta n &=&  {1\over \pi\xi^2} \int_1^\infty {dy \over 2 \sqrt{y^2-1}\left(y+\sqrt{y^2-1}\right)}\nonumber\\
&=& {1\over 4\pi\xi^2}\ .
\label{cond-frac}
\eea

The dimensionless  parameter which controls the higher order terms in the saddle point expansion of Eq.
 (\ref{Zbog}), is then $\delta n/n_0=1/(4\pi n_0\xi^2)$,
which serves as an effective 'quantum disorder' coupling constant. 

For a uniform condensate,  the leading order structure factor is given by
\be
S^{\rm Bog} =  {\epsilon_\bfq\over E_\bfq}+ \cO(n_0\xi^2)^{-1}
\label{Sbog}
\ee
where the order $1/(n_0\xi^2)$ corrections go beyond the  leading order approximation and hence ignored.
An artifact of the wavefunction $\rmPhi $, in (\ref{varphi-Bog}) is that it produces a  
non-vanishing contribution at zero momentum (i.e. an infinite compressibility) from  the fluctuations correlator $\langle\rmPhi  |(\eta)^4|\rmPhi \rangle $.
These are cancelled by self energy corrections due to cubic interactions  $g\varphi^* \eta^* (\eta)^2$, which we will not calculate here.

A quantum phase transition into another zero temperature phase (\eg\ a solid)
may be expected when $\delta n\gtwid n_0$. Here we shall not explore the
strong coupling regime.

\subsection{Full Vortex Bogoliubov Theory}
\begin{figure}[htb]
\begin{center}
\includegraphics[width=9cm,angle=0]{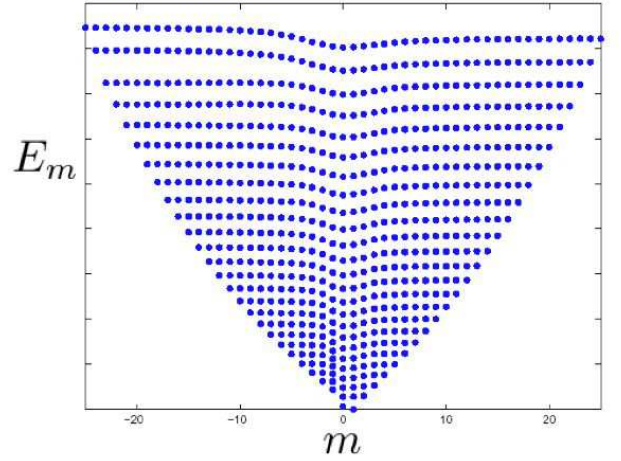}
\caption{The exact Bogoliubov spectrum for a vortex pair configuration on the
sphere, as a function of azimuthal quantum number $m$.  The relevant parameters are
$\xi/R=0.1$, and  a stabilizing rotation  frequency of $\omega_{\rm rot}=
1.875\,\hbar / m R^2$. Note the non-degenerate zero mode at $m=1$. Note the vortices induced distortion of the spectrum
around $m\approx 0$  in a broad energy range.}
\label{fig:bec-spectrum}
\end{center}
\end{figure}

In Fig. \ref{fig:bec-spectrum}, the numerical fluctuation spectrum about
 a vortex pair configuration on a sphere is plotted as a function of angular momentum.

Eq. (\ref{GP}) has a  zero mode  \cite{CastinLH} ($E_n=0$) corresponding to a global U(1)  phase transformation $\phi\to\phi+\delta$, given by
\bea
U_0 (\bfx)=A\,{\widetilde\varphi}(\bfx) &\quad,\quad&  V_0(\bfx)=A^*\,
{\widetilde\varphi}^*(\bfx)\nonumber\\
\int\!d^2\!x\,\big( |U_0|^2-|V_0|^2 \big)&=&0\ .
\label{zeromode}
\eea
This (unnormalizable) zero mode, which enforces charge conservation,
is henceforth excluded from our numerical spectra.

The Bogoliubov eigenoperators (quasiparticles) are given by
\be
a_n = \int\!\! d^2\!x\, \Big( U^*_n(\bfx)\, \eta(\bfx)-   V_n^*(\bfx)\,\eta^*(\bfx)\Big).
\label{Bogqp}
\ee

\begin{figure}[htb]
\begin{center}
\includegraphics[width=10cm,angle=0]{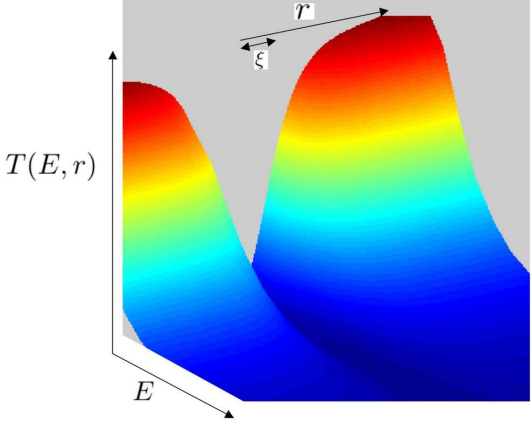}
\caption{The Bogoliubov tunneling density of states of a BEC vortex (in arbitrary units)
 as a function of excitation energy $E$ and radial distance  from the vortex center $r$. $\xi$ is the coherence length.}
\label{fig:bec-tdos}
\end{center}
\end{figure}

To visualize the Bogoliubov fluctuations near a vortex, we plot in  Figure (\ref{fig:bec-tdos}) the tunneling density of states as defined by 
\be
T(E,\bfx)=\pi \sum_n |V_n(\bfx)|^2 \delta (E-E_n),
\ee
which might prove interesting  for inelastic scattering experiments of rotating condensates
 \cite{Steinh}.

The Bogoliubov corrected ground state $\ket{\rmPhi }$ is  the $a$-vacuum, and thus
satisfies $a_n\ket{\rmPhi }=0$. This may be written in terms of the original bosons as
\bea
 \big| \rmPhi  \rangle &=&\cN\exp\big(\half  \psi^\dagger\,Q\,\psi^\dagger\big)\,
\exp\big(f  \psi^\dagger\big)\, \big|0\big\rangle\nonumber\\
 f(\bfx)  & \equiv &  {\widetilde\varphi}(\bfx) -  \int\!\!d^2 x' \,{\widetilde\varphi}^*(\bfx') 
\, Q(\bfx',\bfx)  
\label{varphi-Bog1}
\eea
Integration over coordinates is implied in (\ref{varphi-Bog1}), and the normalization factor
is $\cN=\braket{\rmPhi }{\rmPhi }^{-1/2}$.
The pair  operator $Q$ is 
\be
Q(\bfx,\bfx') =  \langle \bfx| (U^\dagger)^{-1}  V^\dagger|\bfx'\rangle =Q(\bfx',\bfx),
\ee
where elimination of the zero mode (\ref{zeromode}) in $U$ and $V$ is implicitly assumed.

We next compute the overlap of a vortex state with a displaced vortex.
For  two vortex states (\ref{varphi-Bog})  centered at $\bfX_i$, $i$=1,2.
the operators  $Q_i(\bfx,\bfx')$ are distinct and not translationally invariant. 
The magnitude of the wavefunction overlap for Bogoliubov-corrected vortex
states is  
\bea
\big|\braket{\rmPhi_1}{\rmPhi_2}\big| &=&  \exp\left( -W_0-W_1\right)\nonumber\\
W_0&=& W_0^{(12)}-W_0^{(11)}\quad , \quad W_1=-\ln\left|{D_{12}\over D_{11}}\right|
\nonumber\\
{W}_0^{(ij)}&=&   \half (f_i^* f_j )
\left( \begin{array}{cc}
1& -Q_j\\
-Q_i^* & 1
\end{array}
\right)^{-1} 
\left(
\begin{array}{c}
f_j\\
f^*_i
\end{array}
\right)  \nonumber\\
{D}_{ij}&=& {\rm det}^{-1/2}\left( 1- Q^*_i Q_j \right)  \ .
\label{unnorm-Bose-exp}
\eea 
We note that the leading order  $W_0$  is proportional to the condensate density $n_0$, 
while the fluctuation correction $W_1$ depends on $d,\xi, R$, but not on $n_0$. $W_1$ turns out to diverge logarithmically with system size. 
However, this need not concern us here, since we know that at the same order,  the  structure factor of $\rmPhi $  is incomplete, as discussed
after Eq.(\ref{Sbog}).

We now show that in contrast to the 
coherent state overlap result (\ref{MF-IR}), there is no logarithmic divergence, to leading order, in the exponent $W_0(d,R)$.
This can be shown analytically by separating contributions of the core and the far field regions, as depicted in Fig. \ref{fig:vor-cs},
\be
 {W}^0 \sim  \int\limits_{\rm core}\!\! d^2\!x\,  w_{\rm core} (\bfx) +
 \sum_l A_l\, w_{\rm far}(\bfx_l)\ .
 \label{sep}
\ee
The far field integral is  approximated by a sum over constant phase domains, at radii
$r_l\gg\xi$,  and of areas $A_l << r_l^2$, corresponding to the condensate field
$\varphi_{i,l}=\sqrt{n_0}\,\exp(i\phi_{i,l})$.  
We  use the uniform solutions (\ref{uniform}) in each such domain,  assuming that the
block sizes are large enough:  $A_x \gg\xi ^2$. Thus,
\bea
Q_{i,l}&=&  - { e^{2i \phi_{i,l}} \over A_l} \sum_\bfk Q_\bfk\,  e^{i\bfk\cdot(\bfx-\bfx')}
\nonumber\\
&=&  -  e^{i 2 \phi_{i,l}}\Big(\delta(\bfx-\bfx') +\cO(\xi/r_l)\Big)\ ,
\label{uniformQ}
\eea
which, by (\ref{varphi-Bog}), yields in each block $l$, and vortex configurations $i=1,2$, the constant functions
\bea
f_{i,l}(\bfx)&=& \sqrt{n_0}\, \left( e^{i\phi_{i,l}} - e^{-i\phi_{i,l}}\int_l d^2\!x' \,
Q_{i,l}(\bfx,\bfx') \right)  \nonumber\\
&=& 2 \sqrt{n_0} \, e^{i\phi_{i,l}}+\cO(\xi/R) .
\label{uniformf}
\eea
Using   (\ref{unnorm-Bose-exp}) we obtain to zeroth order in $\xi/r_l$ the result
\begin{widetext}
\be 
w_{\rm far}(\bfx_l)\simeq  
n_0\,{\rm Re} \left\{ (e^{-i\phi_{1}}   e^{i\phi_{2}} )
{ \left( \begin{array}{cc}
1&-e^{i 2 \phi_{2} } \\
-e^{-i 2 \phi_{1} }  & 1
\end{array}
\right)   \over  1-e^{ i2(\phi_{2}-\phi_{1})} }
\left(
\begin{array}{c}
e^{i\phi_{2}} \\
e^{-i\phi_{1}}
\end{array}
\right)  - \left( \phi_2 \Leftrightarrow \phi_1\right)\right\}  =  
4 n_0\,{\rm Re} \left( { e^{i(\phi_2-\phi_1)} \over  1+e^{i(\phi_2-\phi_1)} }-\half\right) =0\ ,
\label{Bose-exp}
\ee 
\end{widetext}
where we have suppressed the block index $l$.
Unlike in the coherent states overlap exponent which exhibits a logarithmic divergence (\ref{MF-IR}), 
$W_0$ is perfectly finite in the large system limit 
\be
\sum_l A_l w^{\rm far}_l = \int\limits_{r_0}^\infty\!\! d^2\! r\, (\bfd\cdot\bnabla \phi)^2
\times \cO\left({\xi/ r}  \right)<\infty
\label{sep2}
\ee  
which is in agreement with NAT (\ref{nat}) and (\ref{Sbog}).

 \subsection{Vortex Overlap in BCS States}
\label{app:BdG}
It is perhaps  little appreciated in the literature that  vortex wavefunctions of  BCS  theory of superconductors,
as derived by leading order Bogoliubov de-Gennes equations, 
suffer  from the same {\em overlap orthogonality catastrophe} and infinite compressibility as the Bose coherent states described above. 

We shall demonstrate this point by considering a `generic' BCS superconductor, 
which is described by a microscopic Hamiltonian of attractively  interacting fermions:
\be 
\cH  =  \!\int \!\!d^2\! x\,\Big\{ \psi^\dagger_s\, (K+V) \,\psi\nd_s   -{g\over 2}
\psi^\dagger_s  \psi\yd_{s'} \psi\nd_{s'}  \psi\nd_{s}\Big\}  \nonumber\\
\label{Hfermi}    \ee 
where $\psi_s(\bfx)$ creates an electron of mass $m$, charge $q$ and spin
$s=\uparrow,\downarrow$ at position $\bfx$,
and the operators of kinetic ($K$) and potential ($V$) energy are the same as defined for bosons in  (\ref{K},\ref{GPmodel}).
Summation over repeated spin indices $s,s'$ is assumed.  The complex superconducting order parameter is $\varphi(\bfx)=\langle \psi_\uar(\bfx)\,\psi_\dar(\bfx)
\rangle$.  At long distances from the edges or vortex cores, $\varphi$
minimizes the Ginzburg-Landau energy, i.e. it satisfies Gross Pitaevskii equation  (\ref{GP})  with pair mass $2m$ and charge $q=2e$.
Its magnitude is given by the BCS gap parameter, \ie\ $\Delta=g\big|\varphi\big|$.
As for the BEC, in the presence of a weak magnetic field,
a  quantized vortex solution minimizes the mean field energy, and its core size is given by
the coherence length $\xi  \simeq  \hbar v_\rmF/ \pi\Delta_0$,
where $v_\rmF$ is the Fermi velocity \cite{tinkham}.

 The Bogoliubov de-Gennes (BdG) equations for the superconductor are
 \bea
 H_0 U_n(\bfx) +
g\varphi V_n(\bfx) &=& E_n  U_n(\bfx)\nonumber\\
 g\varphi^*  V_n(\bfx) - H_0^* U_n(\bfx) &=&  E_n V_n(\bfx)\ , \nonumber
\label{FermiBdG}
\eea
where $H_0=K-\mu$ and $\mu=\hbar^2 k_\rmF^2/2m$,
along with  the  self consistency condition,
\be
\sum_n U_n (\bfx) V_n^* (\bfx) =  \varphi (\bfx)\ .
\label{self-cons}
\ee
The self consistency determines the detailed profile of $|\varphi(\bfr)|$ in the core region. 
\begin{figure}[htb]
\begin{center}
\includegraphics[width=9cm,angle=0]{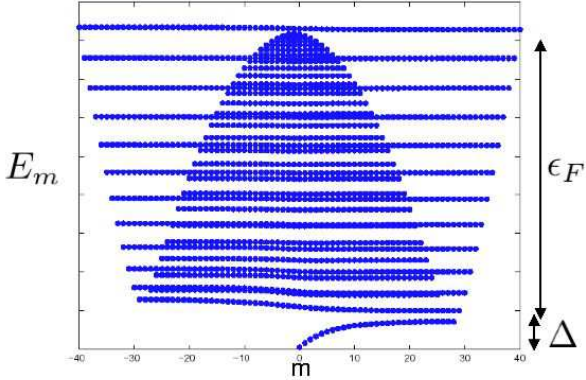}
\caption{The   Bogoliubov-de Gennes spectrum for a vortex pair configuration on the sphere,
as a function of azimuthal quantum number $m$.  The relevant parameters are the BCS gap
$\Delta$, and the Fermi energy $\epsilon_\rmF$. Note the branch of (doubly degenerate) 
low energy Caroli-de Gennes-Matricon core states at positive angular momenta.}
\label{fig:bdg-spectrum}
\end{center}
\end{figure}

Using the solutions of (\ref{FermiBdG}) the vortex ground state is given by
\bea
\ket{\rmPhi }  &=&  \cN \exp\left(   \psi_\uparrow^\dagger\, Q\,
\psi_\downarrow^\dagger    \right)  \ket{0} \nonumber\\
Q &=&  \expect{\bfx}{(U^\dagger)^{-1}  V^\dagger}{\bfx'}\ ,
\label{varphi-BdG}
\eea
where $\cN$ is the normalization.
The normalized overlap of two BCS vortex states displaced by $\bfd$ is
given by
\be
 e^{-W(d)}=\left| \det\left(  U_1^\dagger U_2+
V_1^\dagger V_2 \right)\right|\ .
\label{ov-fer}
\ee

We first calculate the  far field contribution, and see that it exhibits from a similar logarithmic divergence as the boson mean field wavefunctions.  
Asymptotically far  from
the  cores, whose sizes  are of  given by $\xi$, one can  diagonalize (\ref{FermiBdG}) using
a constant order parameter $\varphi_i=\Delta\, e^{i\phi}$. The solution of
(\ref{FermiBdG}) yields
\bea
U^2_\bfk &=&  \half \left( 1+ {\zeta_\bfk\over \sqrt{\zeta_\bfk^2+\Delta^2}}\right)\\
V^2_{\bfk} &=&\half \left( 1 -  {\zeta_\bfk\over \sqrt{\zeta_\bfk^2+\Delta^2}} \right) e^{i2\phi}
\eea
with $\zeta_\bfk=\hbar^2 (k^2-k_\rmF^2) /2m$.
Factorizing the determinant into  blocks, as shown in  Fig. \ref{fig:vor-cs}, we obtain
\bea
W&=& W_{\rm core}+
 \sum_{r_l>r_0} A_l\, w^{\rm far}_l \nonumber\\
w_l^{\rm far}&\simeq &  \big[(  1-\cos\big(\phi_{1,l}-\phi_{2,l}\big)\big]
 \int\!\! {d^2\!k\over (2\pi)^2}\, U^2_{\bfk}
V^2_{\bfk  } \nonumber\\
W&\simeq&\half\pi\,n_{\rm eff}\, d^2 \ln(R/r_0)  +W_{\rm core}\ ,
\label{ov-fer-far}
\eea
where $n_{\rm eff}=k_\rmF/4\pi\xi < n_\rme$.

The divergence of the first term with $\log(R)$ arises from the far-field contributions. Again, it is an artifact of the BCS wavefunction, which is a mean field state where the condensate phase does
not fluctuate.
As for Bose coherent states, the BCS wavefunction also has a non vanishing structure factor at zero momentum, and thus an infinite compressibility at zero temperature.
This divergence is cancelled by including  (RPA) phase fluctuations, which restores a finite zero temperature compressibility  given by the density of states at the Fermi energy.

\begin{figure}[htb]
\begin{center}
\includegraphics[width=9cm,angle=0]{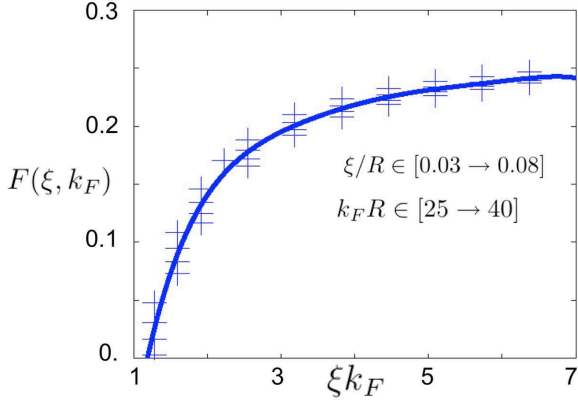}
\caption{The function $F(k_F,\xi)$ defined in Eq.(\ref{WBCS}), which yields the coefficient of the core
contribution to the BCS vortex overlap on the sphere. $k_F$ is the Fermi wavevector and $\xi$ is Pippard's coherence length.
The numerical data in the specified ranges of
 $k_F R$ and $\xi/R$, collapse  to a function of the product $\xi k_F$.}
\label{fig:bdg-asymp}
\end{center}
\end{figure}

Here we are interested in the core contribution $W_{\rm core}$ of Eq. (\ref{ov-fer-far}).  This requires a full diagonalization of the
BdG equations.  For the antipodal vortex pair on  the sphere, the details of the computation are found in 
Appendix (\ref{app:calcs}). 
A fit to the  numerically obtained values of  $W$ in Eq. (\ref{ov-fer}) 
yields the asymptotic  expressions at large $R/d$:
\be 
W^{\rm BCS}  \simeq   d^2 \left( {k_F\over 8\xi}\ln(R/\xi) +   k_F^2 F(k_F\xi)\right)+ \cO(d^4),
\label{WBCS}
\ee
where $F$ is a dimensionless function of order 0.2, of  the  scaling variable $(k_F\xi)$ as demonstrated by the collapse of the 
numerical data for $W$   in Fig. (\ref{fig:bdg-asymp}).

The first $R$-dependent term is the diverging far field contributions discussed earlier, which is an artifact of the infinite compressibility of the BCS wavefunction.

The second term is the core contribution which is in fact very large: it goes as  $(k_f d)^2$. Thus we find that the core contribution
effectively suppresses  energy conserving tunneling between sites separated by more than a few  Fermi
wavelengths. In the regime of large coherence length relative to the  mean free path, vortices could be dissipatively dragged by the bias current 
by exciting low energy core states  \cite{corestates}). This is essence the source of friction in Bardeen-Stephen flux creep theory \cite{BS}. 
This 'frictional' motion involves  impurity scattering inside  the  vortex cores  \cite{CT-2,CT-3}, which we do not treat in this paper.

\section{Calculations on the Sphere}
\label{app:calcs}
An antipodal vortex pair field can be expanded as
\be
{\varphi}(\theta,\phi) = \sum_{l=1}^\infty{\widetilde\varphi}_l\,
Y_{l,1}(\theta,\phi)  
\label{vor-sphere}
\ee
where $Y_{lm}(\theta,\phi)$ are normalized spherical harmonics.
Eq. (\ref{vor-sphere})  is an eigenfunction of $L^z=-i\hbar\,\pz_\phi$. 
The coherence length $\xi$ determines the core sizes at the north and south
poles, and thus the the decay rate of the ${\widetilde\varphi}_l$ with $l$.

The angular momentum representation, with a cut-off at $l_{\rm max}\gg (R/\xi)$, 
reduces the Bogoliubov equations (\ref{BoseBdG},\ref{FermiBdG}) to  finite
matrix diagonalizations.  We can also  translate the  vortex wavefunctions
using  SU(2) rotation $D^l_{mm'}$ matrices.

\subsection{BEC with Antipodal Vortex Pair}
For the BEC vortex pair,  some straightforward but  tedious algebra, can bring the overlap exponent
in Eq. (\ref{unnorm-Bose-exp})  into a  computationally
convenient form
\bea
W_0 &=& {\rm Re}\, \Bigg[ (\varphi^*_1\,U_1-\varphi_1\,V_1)S^{-1}_{12} \nonumber\\
&&\qquad \times\, (U^\dagger_2\,
(\varphi_2-\varphi_1)-V^\dagger_2\,(\varphi^*_2-\varphi^*_1)) \Bigg] \nonumber\\
W_1&=& \half \ln  {\rm det} [S_{12}] \nonumber\\
S&=&  U^\dagger_1\,U_2 - V^\dagger_1\,V_2 \nonumber\\
\label{Bose-overlap-expl}
\eea

For the vortex pair field ${\widetilde\varphi}$ given by  (\ref{vor-sphere}) 
the  BdG equation possesses  axial symmetry which
allows  $m=\bar{m}_n$ to be a good quantum number, and the BdG eigenvectors are 
  \bea
 U^n_{l,m}&=& U^n_{l} \delta_{m, {m}_n},\nonumber\\
 V^n_{l,m} &=&  V^n_{l'} \delta_{m , {m}_n-2}\nonumber\\
 \delta_{nn'} &=&\sum_l U^n_l U^{n'}_l - V^n_l V^{n'}_l
 \eea
for these coefficients is
\bea
E_n \left( \begin{array}{c}
 U^n \\
-V^n \end{array} \right)&=&  \left(\begin{array}{cc}
H^N    & A  \\
-A^t  &  - H^N 
 \end{array} \right) 
\left( \begin{array}{c}
 U^n \\
-V^n \end{array} \right) \\
H^N(m)&=&    \frac{\hbar^2}{2m R^2} l (l +1) \delta_{ll'} +V_{ll',m} \nonumber\\
V_{ll',m} &=& \expect{l,m}{ V_{\rm pin}(\theta)+2g|{\widetilde\varphi}|^2-\mu} {l',m}\nonumber\\
A_{ll'}(m) &=& g  \expect{l,m}{{\widetilde\varphi}^2}{l',m-2}
\label{BdGBS}
\eea
where (\ref{3-j}) can be used for the precise numerical evaluation  of the matrix elements.


\subsection{Boson Ground State Overlap}
Having determined $U^n_l,V^n_l$, and recognizing that the transformation between $\varphi_0, U_0.V_0 $ and $\varphi_1,U_1.V_1$
simply involves an O(3) rotation of the $z$ axis by an  angle $\theta$, the overlap (\ref{Bose-overlap-expl}) is given by
\bea
\braket{\rmPhi_1 }{\rmPhi_2} &=& {\rm det}^{-1/2} [S]\, \exp
\big[ \vec{\varphi}^\rmt  I  S^{-1}  J    \vec{\varphi} \big]
\nonumber\\
S_{nn'}&= & \sum_{l} \left(U^n_{l} D^l_{m_n,m_{n'}} U^{n'}_{l} - V^n_{l } D^l_{m_n,m_{n'}}  V^{n'}_{l}  
\right) \nonumber\\
I_{l,n}&=& \left(U^{n}_{l} + V^n_{l}\right) \delta_{ {m}_n,1}\nonumber\\
J_{n ,l } &=&   U^{n}_{l} \left( \delta_{ {m}_n,1} - D^{l}_{ {m}_n,1} \right) 
+ V^{n}_{l} \left( \delta_{ {m}_n ,1} - D^{l}_{ {m}_n -2,-1} \right)\nonumber\\
\vec{\varphi}_l &=& \!\!\int\!\! d\rmOmega\, Y_{l,1}^*(\theta,\phi)\, \varphi(\theta,\phi)\ .
\label{overlap-BS}
\eea
 It is easy to verify that for the limit $U^n_l=\delta_{n,l}, V^n_l=0$, (\ref{overlap-BS}) reduces to the result for free particles.

\subsection{Fermions on a Sphere}
A polar  vortex pair field  described by Eq. (\ref{vor-sphere}), defines the pairing order parameter
\be
\Delta =  {\widetilde\varphi}(\theta,\phi)
\ee
by normalizing it such that  $\sqrt{n_0}=\Delta_0$.  In the spherical harmonic basis
$\ket{l,m}$ the Hamiltonian (\ref{FermiBdG}) simplifies greatly. 
The Laplacian is proportional to the diagonal operator $\bfL^2$, and non diagonal
matrix elements of functions $F(\theta,\phi)$ can be computed using $3j$ Racah
coefficients  \cite{edmonds}:
\bea
&&\expect{l,m}{F}{l',m-M}\nonumber\\
&&\qquad\qquad= \sum_{L,M}^\infty F_{LM} 
 \left[ {(2l+1)(2L+1)(2l'+1)\over 4\pi}\right]^{1/2}\nonumber\\
&& \qquad\qquad\qquad\times(-1)^m \left(\begin{array}{ccc}
l&L&l'\\ -m& M & m-M \end{array} \right)  \left(\begin{array}{ccc}
l&L&l'\\ 0& 0 & 0 \end{array} \right)\nonumber\\
 && F_{LM} \equiv \! \int\!\! d\rmOmega\, Y^*_{L,M}(\rmOmega)\, F(\rmOmega)
\label{3-j}
\eea

Due to axial symmetry, $m$ is a good quantum number, which is to say that  a function 
 $ m_n$ is defined such that
 \bea
 U^n_{l,m}&=& U^n_{l} \delta_{m,m_n},\nonumber\\
 V^n_{l,m} &=& V^n_{l'} \delta_{m', {m}_n-1}
 \eea
The matrix BdG equation for these coefficients is
\bea 
E_n \left( \begin{array}{c}
 U^n  \\
V^n  \end{array} \right) &=& 
\left(\begin{array}{cc}
H^N  & A  \\
A^\dagger  &  - H^N 
 \end{array} \right) 
\left( \begin{array}{c}
 U^n  \\
V^n  \end{array} \right) \nonumber\\
H^N  &=&  \left(\frac{\hbar^2}{2mR^2}l(l+1) - \ve_{F} \right) \delta_{ll'}  \nonumber\\
A_{ll'}(m) &= &   \expect{l,m}{{\widetilde\varphi}}{l',m -1}
\eea

The overlap of two vortex pair states relatively rotated by $\theta$ is  
 \bea
\braket{\rmPhi_1 }{\rmPhi_2} &=&\det_{nn'} \bigg[  \sum_{l} 
\Big(U^n_{l} D^l_{ {m}_n, {m}_{n'} }  U^{n'}_{l} \\
&&\qquad\qquad+ V^{n}_{l} D^l_{ {m}_n-1, {m}_{n'}-1} V^{n'}_{l}  \Big) \bigg]\nonumber
\eea
where $D^l_{mm'}(\theta)$ is the orthogonal rotation matrix.
 

\section{The Magnus Action}
\label{app:Magnus}
The calculation of the Berry  phase for the motion of a vortex wavefunction can be done using the method of Arovas,
Schrieffer, and Wilczek,  \cite{aro84}, originally for quantum Hall effect quasiparticles, and later
applied {\it mutatis mutandis\/} to superfluid vortices by Haldane and Wu \cite{HW}.
Here we  show how the Berry phase is calculated simply and exactly on a spherical geometry.

The spherical geometry yields many advantages for  vortex
wavefunctions.  The geometric phase of a moving vortex is tricky to evaluate for
superfluids and superconductors on a finite plane, since it sensitively depends on the
boundary conditions.  On the sphere, there are no boundaries to worry about.
The translations are implemented by O(3) rotations, whose generators do not commute.
Hence we can show that the Magnus density of a general many-body wavefunction is
simply given by the expectation value of angular momentum density.  When applied to
an antipodal vortex pair state, this results agrees with Thouless, Ao and Niu's
conclusion  \cite{TAN96} that the Magnus density of a single vortex is just the superfluid
density away from the vortex cores. 

\begin{figure}[htb]
\begin{center}
\includegraphics[height=8cm,width=9cm,angle=0]{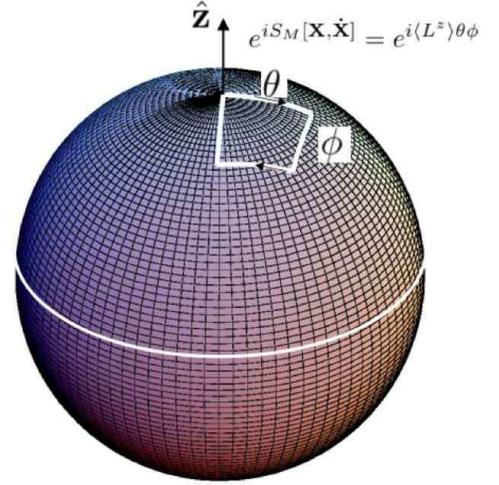}
\caption{The orbit of a vortex pair on the sphere which is used to calculate the vortex Berry phase and Magnus action. Motion of the antipodal vortex 
pair is achieved by 
four successive rotations of the north pole, covering a solid angle of $\theta\times \phi$.}
\label{fig:berry}
\end{center}
\end{figure}

Consider a general  many body wavefunction $|\rmPsi_0\rangle $ defined on a sphere of
radius $R$,  and calculate the Berry phase acquired by an  infinitesimal loop of area 
$A=R\delta\theta\cdot R\delta\phi$ in parameter space, as depicted by Fig. \ref{fig:berry}.

The infinitesimal loop is divided into four segments $\langle i,i+1\rangle$ such that the  loop Berry phase is 
 \be
e^{i S_\rmM/\hbar}\simeq \prod_{i=0}^3\braket{\psi_{i+1}}{\psi_i}\ ,
\label{Magden-def}
\ee
For a general wavefunction, the loop can be defined by a succession of small O(3)
rotations of a vector passing through a special point on the sphere.  For our purpose,
we pick that point to be the vortex center.  For a general wavefunction, however,
the choice of special point is somewhat arbitrary.  Inversely, for a given sequence of small rotations, each point on the sphere executes
a an orbit. The special points on the sphere which characterize the sequence are the two antipodal points whose loops have a maximal area.
(see Fig. \ref{fig:berry}).  
Without loss of generality, we choose to place these points at the north and south poles, and 
and apply the corresponding sequence of rotations:
\bea
\ket{\psi_1} &=& e^{iL_y \delta\theta/\hbar}\,\ket{\psi_0}\nonumber\\
\ket{\psi_2} &=& e^{iL_x \delta\phi/\hbar }\,e^{iL_y \delta\theta/\hbar} 
\,\ket{\psi_0}\nonumber\\
\ket{\psi_3} &=& e^{-iL_y \delta\theta/\hbar}\,e^{iL_x \delta\phi/\hbar }\,
e^{iL_y \delta\theta/\hbar}\,\ket{\psi_0}\nonumber\\
\ket{\psi_4}&=&\ket{\psi_0}
\eea
The overlaps uto quadratic order in $\delta\theta$ and $\delta\phi$ is found to be
 \bea
e^{iS_\rmM/\hbar}\approx 1-i\,\delta\theta\,\delta\phi\,\langle L_z\rangle\big/\hbar + \ldots
 \label{Magden-sp}
\eea
where the expectation values are in the state $\ket{\psi_0}$.

Generalizing to a complete path, we obtain
\be
S_\rmM= - \int\!\!dt\,\expect{\psi_0(t)}{\bfL}{\psi_0(t)}\cdot{d\bfomega\over dt},
\ee
where the vector ${\dot\bfomega}$ is the rate of change of solid angle, instantaneously
directed toward the vortex core.

This Magnus action applies to any wavefunction, regardless of its correlations, 
(\eg\ superfluid, Fermi liquid or solid).  For an antipodal vortex pair in a superfluid, the
angular momentum density is given  by $\hbar$ times the condensate number density
$n_0$, if one assumes that the thermal excitations (normal component) carry no angular
momentum. representing the vortex center coordinates by $(X,Y)$, yields the Magnus Lagrangian 
\be
\cL_{M}= 2\pi \hbar n_0 X {\dot Y}
\label{mag-MF}
\ee
which resembles the effect of a uniform magnetic field in the $z$ direction.

The fluctuations correction of the Bogoliubov corrected wavefunction (\ref{varphi-Bog})  is given by
\be
\cL_{M}= 2\pi \hbar\left( n_0 + \bar{ \delta n}\right) X {\dot Y}
\ee
Interestingly, although $\delta l_z (\bfx)\ne \delta n(\bfx)$, the system averaged 
quantities are found to be numerically equal  $\bar{\delta n}=\bar{\delta l_z}$.

\end{document}

%% file: ldefs.tex
  \font\elevenmib=cmmib10 scaled 1095
  \font\tenmib=cmmib10
  \font\eightmib=cmmib10 scaled 800
  \font\sixmib=cmmib10 scaled 667
  \skewchar\elevenmib='177
  \newfam\mibfam
  \def\mib{\fam\mibfam\tenmib}
  \textfont\mibfam=\tenmib
  \scriptfont\mibfam=\eightmib
  \scriptscriptfont\mibfam=\sixmib
  \mathchardef\alpha="710B
  \mathchardef\beta="710C
  \mathchardef\gamma="710D
  \mathchardef\delta="710E
  \mathchardef\epsilon="710F
  \mathchardef\zeta="7110
  \mathchardef\eta="7111
  \mathchardef\theta="7112
  \mathchardef\kappa="7114
  \mathchardef\lambda="7115
  \mathchardef\mu="7116
  \mathchardef\nu="7117
  \mathchardef\xi="7118
  \mathchardef\pi="7119
  \mathchardef\rho="711A
  \mathchardef\sigma="711B
  \mathchardef\tau="711C
  \mathchardef\phi="711E
  \mathchardef\chi="711F
  \mathchardef\psi="7120
  \mathchardef\omega="7121
  \mathchardef\varepsilon="7122
  \mathchardef\vartheta="7123
  \mathchardef\varrho="7125
  \mathchardef\varphi="7127
    \def\physgreek{
    \mathchardef\Gamma="7100
    \mathchardef\Delta="7101
    \mathchardef\Theta="7102
    \mathchardef\Lambda="7103
    \mathchardef\Xi="7104
    \mathchardef\Pi="7105
    \mathchardef\Sigma="7106
    \mathchardef\Upsilon="7107
    \mathchardef\Phi="7108
    \mathchardef\Psi="7109
    \mathchardef\Omega="710A}

\def\etal{{\it et al.\/}}

\def\ie{{\it i.e.\/}}

\def\eg{{\it e.g.\/}}

\def\ONE{{1\kern-0.60em 1}}
\def\dsl{\raise.15ex\hbox{$/$}\kern-.57em\hbox{$\partial$}}
\def\nsl{\raise.15ex\hbox{$/$}\kern-.57em\hbox{$\nabla$}}
\def\id{\raise.72ex\hbox{$-$}\kern-.85em\hbox{$d$}\,}
\def\gtwid{\,{\raise.3ex\hbox{$>$\kern-.75em\lower1ex\hbox{$\sim$}}}\,}
\def\ltwid{\,{\raise.3ex\hbox{$<$\kern-.75em\lower1ex\hbox{$\sim$}}}\,}
\def\undr{\raise.3ex\hbox{$\sim$\kern-.75em\lower1ex\hbox{$|\vec x|\to\infty$}}}

\def\frac#1#2{{\textstyle{#1 \over #2}}}
\def\half{\frac{1}{2}}

\def\({\left (}
\def\){\right )}

\def\uar{\uparrow}
\def\dar{\downarrow}

\def\cD{{\cal D}}

\def\cH{{\cal H}}

\def\cL{{\cal L}}

\def\cN{{\cal N}}
\def\cO{{\cal O}}
\def\cP{{\cal P}}

\def\cR{{\cal R}}

\def\cV{{\cal V}}

\def\cZ{{\cal Z}}

\def\bfA{{\mib A}}
\def\bfB{{\mib B}}

\def\bfE{{\mib E}}

\def\bfJ{{\mib J}}

\def\bfL{{\mib L}}

\def\bfR{{\mib R}}

\def\bfV{{\mib V}}

\def\bfX{{\mib X}}

\def\bfa{{\mib a}}

\def\bfd{{\mib d}}

\def\bfj{{\mib j}}
\def\bfk{{\mib k}}

\def\bfq{{\mib q}}
\def\bfr{{\mib r}}

\def\bfv{{\mib v}}

\def\bfx{{\mib x}}

\def\bfz{{\mib z}}

\def\ve{\varepsilon}

\def\bfve{{\mib\ve}}

\def\xhi{{\raise.35ex\hbox{$\chi$}}}

\def\bfomega{{\mib\omega}}

\def\rmDelta{{\rm\Delta}}

\def\rmPhi{{\rm\Phi}}
\def\rmPsi{{\rm\Psi}}
\def\rmOmega{{\rm\Omega}}

\def\rmF{{\rm F}}

\def\rmM{{\rm M}}

\def\rme{{\rm e}}

\def\rmt{{\rm t}}

\def\ket#1{{\big| #1 \big\rangle}}

\def\braket#1#2{{\big\langle #1 \big| #2 \big\rangle}}
\def\expect#1#2#3{{\big\langle #1 \big| #2 \big| #3 \big\rangle}}

\def\pz{{\partial}}

\def\nd{^{\vphantom{\dagger}}}

\def\yd{^\dagger}

\def\and{a^{\phantom\dagger}}

\gdef\journal#1, #2, #3, 1#4#5#6{               
    {\sl #1~}{\bf #2}, #3 (1#4#5#6)}            

%% file: Vortextransport-eps.bbl
\begin{references}

\bibitem{and66} P. W. Anderson, {\sl Rev. Mod. Phys.} {\bf 38}, 298 (1966).

\bibitem{tinkham} M. Tinkham, {\sl Introduction to Superconductivity},
(Krieger, NY (1980)).


\bibitem{AHNS80} V. Ambegaokar \etal,, {\sl Phys. Rev. B} {\bf 21}, 1806 (1980).

\bibitem{AK64} P. W. Anderson and Y. B. Kim, Rev. Mod. Phys. {\bf 36}, 39 (1964).

\bibitem{tunn-particle} G. Blatter, V. B. Geshkenbein and V. M. Vinokur, {\sl Phys. Rev. Lett.} {\bf 66} 3297
(1991); P. Ao and D. J. Thouless, {\sl Phys. Rev. Lett.} {\bf 70}, 2158 (1993);
M. J. Stephen, {\sl Phys. Rev. Lett.} {\bf 72}, 1534 (1994).

\bibitem{eck89} U. Eckern and A. Schmid, {\sl Phys. Rev. B} {\bf 39}, 6441 (1989);
U. Eckern and E. B. Sonin, {\sl Phys. Rev. B} {\bf 47}, 505 (1993).
R. Fazio, A. van Otterlo, and G. Sch{\"o}n, {\sl Europhys. Lett.} {\bf 25}, 453 (1994);
R. Iengo and G. Jug, {\sl Phys. Rev. B} {\bf 52}, 7536 (1995).
\bibitem{non-linVI} D. Ephron, A. Yazdani, A. Kapitulnik, and M. R. Beasley, {\sl Phys. Rev. Lett.} 76, 1529-1532 (1996);
 J. B. Majer, J. Peguiron, M. Grifoni, M. Tsuveld and J. E. Mooij,
eprint {\tt cond-mat/0301073}.
\bibitem{Creep-exp} A. J. J. van Dalen,  \etal, {\sl Phys. Rev. B} {\bf 54} 1366, (1996);   F. Tafuri, J.R. Kirley, P.G. meaglia, P. Origani and G. Balestrino, Phys. Rev. Lett. {\bf 92}, 157006  (2004).


\bibitem{mpaf91} M. P. A. Fisher, T.A. Tokuyasu and A.P. Young, {\sl Phys. Rev. Lett.} {\bf 66}, 2931  (1991).

\bibitem{mpaf90} M. P. A. Fisher and D. H. Lee, Phys. Rev. B {\bf 39}, 2756
(1989); M. P. A. Fisher, Phys. Rev. Lett. {\bf 65}, 923 (1990); E. Shimshoni, A. Auerbach and A. Kapitulnik, Phys. Rev. Lett. {\bf 80}, 3352 (1998).

\bibitem{SCI-old} A.F. Hebard and M.A. Paalanen, Phys. Rev. Lett. {\bf 65}, 927  (1990);  
A. Yazdani and A. Kapitulnik, Phys. Rev. Lett. 74, 3037
(1995); D. Ephron, A. Yazdani, A. Kapitulnik, and M. R.
Beasley, Phys. Rev. Lett. 76, 1529 (1996); 
\bibitem{SCI-new} G. Sambandamurthy  \etal, Phys. Rev. Lett. {\bf 92}, 107005 (2004); 
M. A. Steiner, G. Boebinger, and A. Kapitulnik
Phys. Rev. Lett. 94, 107008 (2005).

\bibitem{BoseMetal} D. Dalidovich and P. Phillips, Phys. Rev. B {\bf 64}, 052507 (2001).

\bibitem{vol95} G. Volovik, {\sl JETP Lett.} {\bf 62}, 65 (1995).

\bibitem{NAT94} Q. Niu, P. Ao and D. J. Thouless,
{\sl Phys. Rev. Lett.} {\bf 72}, 1706  (1994);
Q. Niu, P. Ao, and D. J. Thouless, {\sl Phys. Rev. Lett.} {\bf 75}, 975 (1995)[C].


\bibitem{duan} J. Duan, {\sl Phys. Rev. B} {\bf 48}, 333 (1993);
{\it ibid.} {\bf 49}, 12381 (1994); J.-M. Duan,  {\sl Phys. Rev. Lett.} {\bf  75}, 974 (1995) [C];

\bibitem{simanek} E. Simanek, 'Inhomogeneous Superconductors', (Oxford University press, 1994)

 \bibitem{aro97} D. P. Arovas and J. A. Freire, {\sl Phys. Rev. B} {\bf 55}, 1068 (1997).



\bibitem{Sonin} E. B. Sonin, {\sl Sov. Phys. JETP} {\bf 37}, 494 (1974);
({\sl Zh. Eksp. Teor. Fiz.} {\bf 64}, 970 (1973)); E. B. Sonin, {\sl Physica B} {\bf 210},
234 (1995).

\bibitem{assaQH} This approach has been used to calculate tunneling rates of fractionally charged quasiparticles
in the Quantum Hall phase:  A. Auerbach,  Phys. Rev. Lett. {\bf 80}, 817  (1998); 
E. Shopen, Y. Gefen, Y. Meir, cond-mat/0503608.


\bibitem{mizel} A. Mizel, cond-mat/0107530.

\bibitem{Uemura} Y. J. Uemura \etal, Phys. Rev. Lett {\bf 62} 2317 (1989). 

\bibitem{AHL} V. Ambegaokar. B.I. Halperin and J.S. Langer, {\sl Phys. Rev. B} {\bf 4}, 2612 (1971);

\bibitem{QHI-theory} O. Viehweger and K.B. Efetov, J. Phys. Condens. Matter {\bf 2}, 7049 (1990);
S. C. Zhang, S. Kivelson, and D. H. Lee, Phys. Rev. Lett. {\bf 69},
1252 (1992); Imry, Phys. Rev. Lett. {bf 71}1868 (1993); L. Zheng and H. A. Fertig,
Phys. Rev. Lett. 73, 878 (1994); O. Entin-Wohlman, A. G. Aronov, Y. Levinson, and Y. Imry,
Phys. Rev. Lett. {\bf 75}, 4094 (1995).

\bibitem{QHI-Shahar} D. Shahar, D. C. Tsui, M. Shayegan, E. Shimshoni, and S. L.
Sondhi, Science {\bf 274}, 589 (1996); M. Hilke, D. Shahar, S.H. Song,
D.C. Tsui, Y.H. Xie and D. Monroe, Nature {\bf 395}, 675 (1998).


\bibitem{SA} E. Shimshoni and A. Auerbach, Phys. Rev. B {\bf 55}, 9817
(1997); E. Shimshoni, A. Auerbach and A. Kapitulnik, Phys. Rev. Lett. {\bf 80}, 3352  (1998).
 
 \bibitem{TAN96} D. J. Thouless, Ping Ao, and Qian Niu,
{\sl Phys. Rev. Lett.} {\bf 76}, 3758 (1996).

 
\bibitem{HL} A standard example of Heitler-London approach can be found in 
D.C. Mattis, `Theory of Magnetism I', (Springer-Verlag, 1988) Ch 2.2.

\bibitem{Comm-HL} Caution must be exercised using Heitler-London method. Since both symmetric and antisymmetric energies
are variational, their difference  is neither an upper nor a lower bound on the exact tunnel splitting.

\bibitem{OF} L. Onsager, Nuovo Cimento Suppl. {\bf  6}, 249 (1949); R.P. Feynamn and M. Cohen, Phys. Rev. {\bf 102}, 1189 (1956).

\bibitem{PS} L. P. Pitaevskii and  S. Stringari,  {\sl  'Bose-Einstein Condensation'} (Clarendon Press, 2003).

\bibitem{Thanksto} We thank Ehud Altman and Anatoly Polkovnikov for alerting us to the spurious Anderson Overlap  Catastrophe associated with coherent states.

\bibitem{Gross1}E. P. Gross, {\sl Nuovo Cimento} {\bf 20}, 454 (1961).

\bibitem{Pit1}L. P. Pitaevskii, {\sl Zh. Eksp. Teor. Fys.} {\bf
40}, 646 ; [{\sl Sov. Phys. JETP}, {\bf 13}, 451] (1961).

\bibitem{Fetter} For example see: A. L. Fetter and A. S. Svidzinsky, {\sl J. Phys. Cond. Matt}
{\bf 13}, R135 (2001); S. Ghosh, {\sl Phase Transitions} {\bf 77}, 625 (2004).

 \bibitem{book1} C. Pethick and H. Smith, {\it Bose-Einstein Condensation in Dilute
Gases} (Cambridge University Press, Cambridge (2002)), Ch. 9.


\bibitem{timescale} The relevant time-scale for tunneling, is the time to move in the inverted  potential barrier
 for the instanton action, which is of the order of $\hbar/\bar{\bf V}$.

\bibitem{hopping} T. Holstein, Ann. Phys. (NY) {\bf 8}, 343 (1959);
 A. Miller and E. Abrahams, Phys. Rev. {\bf 120}, 745 (1960). 
 G. D.  Mahan, ''Many Particle Physics'', (Plenum, NY 1986).

\bibitem{Mott} N.F. Mott, Phil. Mag. {\bf 19}, 835 (1969).


\bibitem{vrh-Sh} B.I. Shklovskii, Zh. Esk. Teor. Fiz. {\bf 36}. 43 (1982),   JETP Lett. {\bf 36}, 51 (1982).
\bibitem{Fogler} M.M. Fogler, A. Yu. Dobin and B.I. Shklovskii,  {\sl Phys. Rev. B} {\bf 57}, 4614 (1998).


\bibitem{sxy-omega} The two limits do not necessarily commute, as known for Anderson insulators
where quantum interference causes localization. Here we are interested
in a strongly dephased system at finite temperatures. We thank Misha Fogler for a clarifying discussion.


\bibitem{Hol-Hall} T. Holstein, Phys. Rev. {\bf 124}, 1329 (1961).


\bibitem{QHI-exp} V. J. Goldman, M. Shayegan, and D. C. Tsui, Phys. Rev. Lett. {\bf 61}; 
 R. L. Willett, H. L. Stormer, D. C.
Tsui, L. N. Pfeiffer, K. W. West, and K. W. Baldwin, Phys. Rev.
B {\bf 38}, 7881 (1988).
 
 \bibitem{Hall-class} D. J. Bergman and D. Stroud, Solid State Phys. {\bf 46}, 147 
(1992);  A. M. Dykhne and I. M.
Ruzin, Phys. Rev. B {\bf 50}, 2369 (1994).
 
 
\bibitem{PA} L. Pryadko and A. Auerbach, Phys. Rev. Lett. {\bf 82}, 1253  (1999).

\bibitem{Balents} L. Balents, L. Bartosch, A. Burkov, S. Sachdev, and K. Sengupta, Phys. Rev. B {\bf 71}, 144508 (2005)



\bibitem{vortex-bec} M.R.Mathews, \etal, {\sl Phys. Rev. Lett.} {\bf 84}, 2498, (1999); 
(2001);  K. W. Madison \etal, {\sl 
Phys. Rev. Lett.} {\bf  84}, 806 (2000); J. R. Abo-Shaeer, \etal Science, {\bf 292}, 476 (2001);
E. Hodby,\etal {\sl   Phys. Rev. Lett.} {\bf  88}, 010405 (2002). 

\bibitem{QED3} M. Franz, Z. Tesanovic and O. Vafek, Phys. Rev. B {\bf 66}, 54535 (2002).

\bibitem{PBFM} E. Altman and A. Auerbach, Phys. Rev. B {\bf 65}, 104508 (2002).

\bibitem{hall-hitc} J. Clayhold, N. P. Ong,   Z. Z. Wang, 
 J. M. Tarascon and P. Barboux, Phys. Rev. B {\bf 39}, 7324 (1989);
 H.Y. Hwang, \etal, Phys. Rev. Lett. {\bf 72}, 2636 (1994).
 
\bibitem{bobinger} F. F. Balakirev, \etal, Nature, {\bf 424}, 912 (2003).



\bibitem{Burnett} M. Edwards \etal, {\sl J. Res. Natl. Inst. Stand. Techol.}
 {\bf 101}, 553 (1996); A. Fetter, {\sl Phys. Rev. A} {\bf 53}, 4245 (1996);

\bibitem{CastinLH}  Y. Castin, in {\sl Coherent atomic matter waves},
Lecture Notes of Les Houches Summer School, edited by R. Kaiser, C.
Westbrook, and F. David, EDP Sciences and Springer-Verlag (2001).



\bibitem{Steinh} J. Steinhauer \etal, {\sl Phys. Rev. Lett} {\bf 88}, 120407 (2002);  J. Steinhauer \etal, {\sl Phys. Rev. Lett} {\bf 90}, 060404 (2003).


\bibitem{corestates} C. Caroli, C.; P.G. de Gennes and J. Matricon, 
Phys. Lett. {\bf 9} 307 (1964).


\bibitem{BS} J. Bardeen and  M. J. Stephen, Phys. Rev. {\bf 140}, 1197 (1965).
\bibitem{CT-2} M. V. Feigel'man \etal, {\sl Sov. Phys. JETP} {\bf 5}, 711 (1993)  ({\sl Pis'ma Eksp.
Teor. Fiz.} {\bf 57}, 699 (1993)); G. Blatter \etal, {\sl Rev. Mod. Phys.} {\bf 66}, 1125
(1994);  A. van Otterlo, M. Feigel'man, V. Geshkenbein, and G. Blatter,
{\sl Phys. Rev. Lett.} {\bf 75}, 3736 (1995).

\bibitem{CT-3} N. B. Kopnin, {\sl Rep. Prog. Phys.} {\bf 65}, 1633 (2002).


\bibitem{edmonds} A. R. Edmonds, {\it Angular Momentum in Quantum Mechanics\/}
(Princeton University Press, Princeton, 1996).


\bibitem{aro84} D. P. Arovas, J. R. Schrieffer, and F. Wilczek, {\sl Phys. Rev. Lett.}
{\bf 53}, 722 (1984).
\bibitem{HW} F. D. M. Haldane and Y.-S. Wu, {\it ibid.} {\bf 55}, 2887 (1985).

\end{references}
